\RequirePackage{fix-cm}
\documentclass[natbib,smallextended]{svjour3}       
\smartqed  
\usepackage[dvipsnames]{xcolor}
\usepackage{graphicx}
\usepackage{amssymb,amsmath,mathrsfs}
\usepackage{hyperref}
\usepackage[normalem]{ulem}  
\hypersetup{
    colorlinks,%
    citecolor=blue,%
    linkcolor=blue,%
    urlcolor=blue
}

 \journalname{mon petit journal}
\newcommand{\xraychap}{X-ray Chapter}

\newcommand{\optchap}{Optical Chapter}

\newcommand{\impostchap}{Imposters Chapter}
\newcommand{\hostchap}{Host Galaxies Chapter}

\newcommand{\disrupchap}{Disruption Chapter}
\newcommand{\flowchap}{Formation of the Accretion Flow Chapter} 
\newcommand{\diskchap}{Accretion Disc Chapter}

\newcommand{\emischap}{Emission Mechanisms Chapter}

\newcommand{\binchap}{Binaries Chapter}

\begin{document}

\title{Rates of Stellar Tidal Disruption}


\author{N.C.~Stone \and E.~Vasiliev \and M.~Kesden  \and E.M.~Rossi \and H.B.~Perets \and P.~Amaro-Seoane }


\institute{N.C. Stone \at
              Racah Institute of Physics \\
              The Hebrew University\\
              Jerusalem, 91904 (Israel) \\
              \email{nicholas.stone@mail.huji.ac.il}           
              \and
              E. Vasiliev \at
              Institute of Astronomy\\ 
              University of Cambridge\\ 
              Madingley Rd\\
              Cambridge, CB3 0HA (UK)\\
              \email{eugvas@lpi.ru}
              \and
            M. Kesden \at
              Department of Physics\\
              University of Texas at Dallas\\
              800 W. Campbell Rd\\
              Richardson, TX 75080 (USA)\\
              \email{kesden@utdallas.edu}
           \and
           E.M. Rossi \at
              Leiden Observatory\\
              Leiden University\\
              PO Box 9513\\
              2300 RA, Leiden (the Netherlands)\\
              \email{emr@strw.leidenuniv.nl}
              \and
            H.B. Perets \at
              Department of Physics\\
              Technion, \\
              Haifa, Israel 3200002\\
              \email{hperets@ph.technion.ac.il}
              \and
            P. Amaro-Seoane \at
              Institut de Ci\`encies de l’Espai\\
              Campus UAB, Carrer de Can Magrans\\
              Cerdanyola del Vall\'es, 08193, Barcelona (Spain)\\
              \email{pau@ice.cat}
}

\date{Received: date / Accepted: date}

\maketitle
\newpage

\begin{abstract}
Tidal disruption events occur rarely in any individual galaxy.  Over the last decade, however, time-domain surveys have begun to accumulate statistical samples of these flares.  What dynamical processes are responsible for feeding stars to supermassive black holes?  At what rate are stars tidally disrupted in realistic galactic nuclei?  What may we learn about supermassive black holes and broader astrophysical questions by estimating tidal disruption event rates from observational samples of flares?  These are the questions we aim to address in this Chapter, which summarizes current theoretical knowledge about rates of stellar tidal disruption, and compares theoretical predictions to the current state of observations.
\end{abstract}

\section{Introduction}
\label{sec:intro}
The discovery and characterization of quasars in the 1960s \citep{Schmidt63} was rapidly recognized as evidence for the existence of supermassive black holes (SMBHs) \citep{Salpeter64}.  Shortly thereafter, the possibility of tidal disruption events (TDEs) was proposed by \citet{Wheeler71}, who hypothesized that the tidal disruption of a star by a massive black hole could create a high energy transient by means of the disintegrational Penrose process.  If stars were consumed frequently enough by massive black holes, the resulting accretion of their gas could, perhaps, explain the observed properties of quasars and active galactic nuclei \citep{Hills75}.

By the mid-1970s, the importance of {\it TDE rates} was evident, and rapid theoretical progress was made on understanding the dynamical processes that feed stars to SMBHs.  Early work envisaged this as a diffusive process in the space of orbital energy: an isotropic distribution of stars living around a SMBH might be depleted by diffusion of stars onto more tightly bound orbits \citep{Bahcall&Wolf76}.  In reality, however, the process of diffusion through energy space is slow and self-limiting; subsequent analytic \citep{Frank&Rees76} and numerical \citep{Lightman&Shapiro77} works quickly demonstrated the greater importance of velocity-space {\it anisotropies} in determining TDE rates.  More precisely, the rate of tidal disruptions in a dense star cluster is set by diffusion through angular momentum, rather than energy, space.  The stellar distribution function (DF) drains into the black hole through a ``loss cone,'' named after the analogous phase space region in magnetic mirror fusion reactors \citep{Rosenbluth&Post65}.

In this Chapter, we survey the dynamical physics of dense stellar systems containing massive black holes, focusing particularly on the collisional evolution of stellar DFs in the presence of a loss cone. Our presentation will, by necessity, be rather terse and without proofs. A more detailed treatment of stellar orbits and kinetic theory can be found in several textbooks.  The reader interested in going beyond this chapter may wish to consult chapter 7 of \citet{Binney&Tremaine08} or chapters 5 and 6 of \citet{Merritt13}.

We begin in \S \ref{sec:TDEBasics}, by overviewing the basic physics of stellar tidal disruption.  In \S \ref{sec:losscone}, we present the theoretical picture of the loss cone, describing both the orbital dynamics of individual stars near supermassive black holes, and the ways in which two-body relaxation and collisionless effects cause stellar populations to evolve over time.  We provide both Newtonian and general relativistic treatments of loss cone dynamics, and examine more exotic types of tidal disruptions (e.g. disruption of evolved or binary stars).  Next, we apply these theoretical tools to realistic astrophysical environments.  In \S \ref{sec:applied}, we examine past efforts to estimate TDE rates by building dynamical models of nearby galactic nuclei, emphasizing both the empirically-calibrated event rate predictions and the distributions of event properties.  In \S \ref{sec:comparison}, we compare these estimates to observational inferences of the volumetric stellar disruption rate.  Finally, in \S \ref{sec:implications}, we examine the broader importance of nuclear stellar dynamics, and describe how well-measured TDE rates may determine the uncertain bottom end of the supermassive black hole mass function, probe basic predictions of general relativity, and calibrate rate estimates for extragalactic phenomena such as mHz gravitational wave sources.

\section{On the Proper Care and Feeding of Supermassive Black Holes}
\label{sec:TDEBasics}
Modern observations demonstrate that SMBHs are ubiquitous in the nuclei of sufficiently large galaxies (e.g. \citealt{Kormendy&Richstone95}).  While a minority of SMBHs live in active galactic nuclei (AGN), where they accrete steadily from long-lived and large-scale discs of interstellar gas, the majority of SMBHs are quiescient: accreting at very low rates \citep{Heckman+04}.  In these quiescent galactic nuclei, it is only in the aftermath of a TDE that the SMBH may shine  brightly.  But what variables determine the rate of stellar tidal disruption?  We can break down the controlling variables into two categories: those relevant to the {\it hydrodynamic} process of tidal disruption, and those related to the {\it orbital dynamics} of stars in galactic nuclei.  

The first set of variables -- those governing hydrodynamic stellar disruption -- are explored in greater detail in the \disrupchap, whereas in this Chapter, we focus on the latter set, with the following philosophy.  In quiescent galactic nuclei, hydrodynamic forces are negligible and the orbits of stars will be shaped by (i) the smooth background gravitational potential of the SMBH and the stellar population; (ii) discrete, pair-wise scatterings with other stellar-mass objects; (iii) coherent secular effects arising from correlations between the orbits of multiple stars\footnote{As we see later in \S \ref{sec:aspherical}, factor (iii) is generally unimportant for determining TDE rates.}.  In the continuum limit, the DF of a population of identical stars can be written in position space as $f(\vec{r}, \vec{v})$.  In a sufficiently old galactic nucleus, the DF will exist in a quasi-steady state, but two-body scatterings and collisionless effects will cause it to evolve adiabatically.  We call such an old nucleus ``relaxed,'' in contrast to a nucleus with a younger stellar population that was born far from a quasi-steady state configuration; such a young nucleus would be ``unrelaxed.''

A full knowledge of the DF $f(\vec{r}, \vec{v})$ and the gravitational potential $\Phi(\vec{r})$ would allow us to compute osculating stellar orbits, their temporal evolution, and the rate at which stars pass near the SMBH.  But in order to understand which orbits are doomed to disruption, we must first introduce at least approximate hydrodynamic criteria for the disruption process.  Fortunately, these approximations are reasonably accurate.

Tidal forces increase as one approaches a black hole.  The strength of the tidal field diverges as distance from the singularity $r\to0$, so interior to some critical distance $R_\mathrm{t}$ (the tidal radius, Eq.~\ref{eq:rtidal}), any macroscopic object will be torn apart.
By equating the Newtonian tidal field to the victim star's surface self-gravity, we find that a star of mass $M_\star$ and radius $R_\star$ will be torn apart by tides if its pericentre $R_{\rm p}$ is roughly\footnote{In reality, the exact criterion for tidal disruption of a main sequence star is that $R_{\rm p} < R_{\rm t}/\beta_{\rm crit}$, where $\beta_{\rm crit}\approx 0.95-1.85$ is a dimensionless constant dependent on the central concentration of the star, and can be measured precisely with numerical hydrodynamics simulations (\citealt{Guillochon&RamirezRuiz13, Mainetti+17}; see also the \disrupchap).} 
within a tidal radius,
\begin{equation}  \label{eq:rtidal}
    R_{\rm t} = R_\star \left( \frac{M_\bullet}{M_\star} \right)^{1/3},
\end{equation}
of an SMBH with mass $M_\bullet$.  This Newtonian expression is reasonably accurate when $R_{\rm t} \gg R_{\rm g}$, where the latter is the gravitational radius $R_{\rm g} \equiv G M_\bullet / c^2$.  But, as we shall see in \S \ref{sec:GR}, when $R_{\rm t} \sim R_{\rm g}$, the tidal radius will depend significantly on SMBH spin $\chi_\bullet$, and the orbital inclination $\iota$. 

Until \S \ref{sec:GR}, however, we will treat Eq. \ref{eq:rtidal} as exact.  With this approximation, we can immediately note one (slightly counter-intuitive) feature of tidal disruption.  Any SMBH will have an event horizon comparable in size to its gravitational radius, $R_{\rm g}$.  We may note that $R_{\rm t} \propto M_\bullet^{1/3}$, while $R_{\rm g} \propto M_\bullet^1$.  The weak power-law dependence of $R_{\rm t}$ on $M_\bullet$ means that, for any given stellar type, there is a {\it maximum black hole mass} capable of producing a TDE outside the event horizon.  TDEs from SMBHs above this critical mass, often called the Hills mass \citep{Hills75}, will produce debris wholly swallowed by the horizon, and will fail to produce an electromagnetically luminous flare.  Approximating the event horizon size as $2R_{\rm g}$, we find that the Hills mass is 
\begin{equation}
   M_{\rm H} = M_\star^{-1/2}\left(\frac{c^2R_\star}{2G} \right)^{3/2} \approx 1\times 10^8 M_\odot \left(\frac{R_\star}{R_\odot}\right)^{3/2}\left(\frac{M_\star}{M_\odot}\right)^{-1/2}. \label{eq:MHills}
\end{equation}
Eqs. \ref{eq:rtidal} and \ref{eq:MHills} are fundamentally Newtonian expressions that we generalize to a relativistic context in \S \ref{sec:GR}.  A relativistic treatment is necessary to account for the effects of SMBH spin, which can sizeably alter $M_{\rm H}$ \citep{Beloborodov+92, Kesden12}.

In both the Newtonian and relativistic regimes, it is clear that, so long as TDE rates depend on $R_{\rm t}$ (the exact dependence is non-trivial and will be quantified in \S \ref{sec:losscone}), they will depend on the mass, radius, and (to a weaker extent) internal structure of the star.  These quantities evolve over the lifetime of a star, so that the tidal radius ultimately depends on the star's zero-age main sequence (ZAMS) mass, its metallicity, its age, its spin, and even its binarity (see \S \ref{sec:types}).  In the general relativistic context (appropriate when $R_{\rm p}\sim R_{\rm g}$), SMBH spin $\chi_\bullet$ and spin-orbit misalignment $\iota$ may affect disruption rates as well.

If we care not only about {\it intrinsic} rates of tidal disruption (e.g. per-galaxy rates, volumetric rates, etc.), but rates of TDE detection in current or planned surveys, then we must consider additional properties of these events.  For example, we may define the strength of the TDE with a dimensionless penetration parameter
\begin{equation}  \label{eq:beta}
    \beta = R_{\rm t} / R_{\rm p}.
\end{equation}
When $\beta \approx 1$, we have a relatively mild, grazing disruption; if $\beta \gg 1$, we have a more violent, deeply plunging disruption; if $\beta \lesssim 1$, a partial disruption may ensue.  
The observational properties of a TDE flare may depend in various ways on $\beta$ \citep{Carter&Luminet83, Stone+13, Dai+15}, meaning that distributions of this parameter are an important prediction for event rate calculations.

As this discussion illustrates, TDE rates depend on a large number of variables related to the participating stars and SMBHs themselves.  Fortunately (for the sake of minimizing astrophysical uncertainty), intrinsic TDE rates depend only weakly on stellar properties, through $R_{\rm t}$.  However, the properties of host galaxies, and their SMBHs, matter a great deal more in setting TDE rates.  In the next section, we will compute first-principles rates of tidal disruption in idealized galactic nuclei, providing the theoretical framework to understand how TDE rates vary with host galaxy properties.

\section{Loss-Cone Theory} \label{sec:losscone}
In this section, we overview, from a theoretical perspective, the many ways in which stars can be fed to massive black holes.  As we shall see, the dynamical evolution of stellar orbits from ``safe'' to ``unsafe'' trajectories is most easily visualized and quantified with the concept of a loss cone, which we define in \S \ref{sec:LCBasics}.  We present the physics of loss cones in spherical galactic nuclei, as well as their connections to TDE rates, in \S \ref{sec:relaxation} and \S \ref{sec:aniso}.  The first of these sections is focused on steady-state loss-cone physics, while the second focuses on the time-dependent problem.  We consider more general (aspherical) geometries in \S \ref{sec:aspherical}, which opens up new avenues for loss-cone repopulation.  In \S \ref{sec:GR}, we generalize our treatment of the loss cone from Newtonian to general relativistic gravity.  Finally, in \S \ref{sec:types}, we briefly survey other types of SMBH loss cones, relevant for processes beyond main-sequence stellar disruption.

\subsection{The Loss Cone}
\label{sec:LCBasics}

In Newtonian gravity, a star on a nearly parabolic orbit (eccentricity $e \approx 1$) with pericentre $R_{\rm p}$ will have specific orbital angular momentum $L \approx \sqrt{2 G M_\bullet R_\mathrm{p}}$.  This implies that stars with specific orbital angular momentum below the critical value (or, equivalently, with $\beta > 1$),
\begin{equation}
    L_\mathrm{\rm t} = \sqrt{2 G M_\bullet R_\mathrm{t}},
\end{equation}
will be disrupted when passing within $R_\mathrm{t}$ of the SMBH.  One can define analogous ``loss regions'' in angular momentum space for destructive processes other than tidal disruption, and we will explore these later on, in \S \ref{sec:types}.

We are generally interested in the evolution of the stellar DF in phase space on timescales much longer than the orbital time $T_\mathrm{orb}$, and will therefore describe populations of stars in the space of integrals of motion (``integral space''), rather than the phase space of $\vec{r}$ and $\vec{v}$.  It is convenient to define a new variable $\mathcal R \equiv \big[ L / L_\mathrm{circ}(E) \big]^2 \in [0,1]$, where $L_\mathrm{circ}(E)$ is the specific angular momentum of a circular orbit with the given specific energy $E$; in a Keplerian potential, $\mathcal R = 1-e^2$.  The number $N(E, \mathcal R)$ of stars per unit interval of $E$ and $\mathcal R$ is related to the DF $f(E, \mathcal{R})$ as \citep[Equation~5.166]{Merritt13}
\begin{equation}
N(E, \mathcal R) = 4\pi^2\, T_\mathrm{orb}(E, \mathcal R)\, L_\mathrm{circ}^2(E)\; f(E, \mathcal R).
\end{equation}
Since we are interested mainly in the low-$\mathcal R$ region, it is convenient to ignore the dependence of orbital time on $\mathcal R$ (which is usually weak) and write $N(E, \mathcal R) \approx g(E)\, f(E, \mathcal R)$, where $g(E)$ is the density of states, defined more generally as:
\begin{equation}  \label{eq:g_of_E}
g(E) \equiv 4\pi^2\,L_\mathrm{circ}^2(E) \int_0^1 T_\mathrm{orb}(E, \mathcal R)\, \mathrm{d} \mathcal R.
\end{equation}
The number of stars per unit energy is $\overline N(E) \equiv \int_0^1 N(E, \mathcal R) \; \mathrm d \mathcal R$. An isotropic distribution of stars in velocity corresponds to a uniform distribution in $\mathcal R$: $N(E, \mathcal R) \approx \overline N(E)$. 
\begin{figure}
\includegraphics[width=120mm]{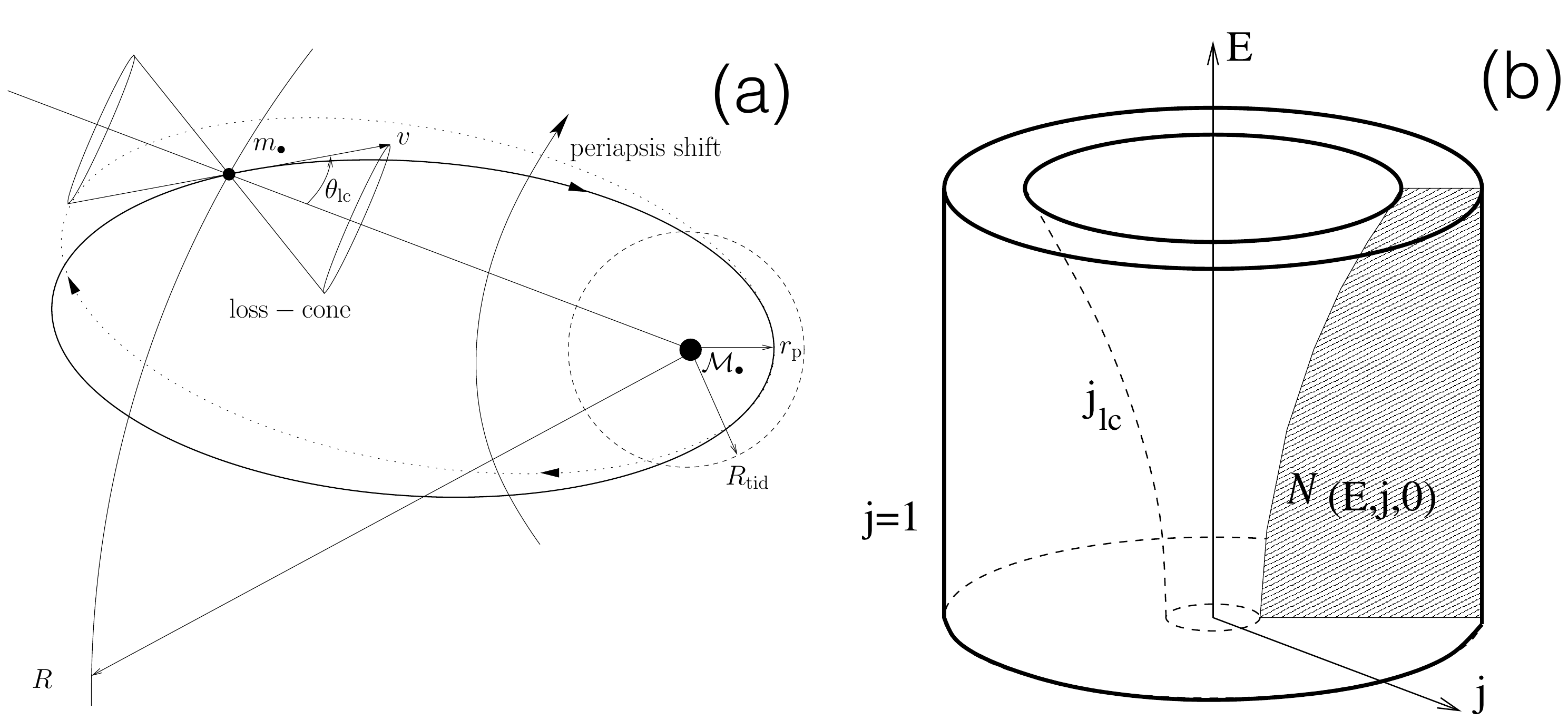}
\caption{Schematic representations of the loss cone in a spherical galactic nucleus.  On the left (panel a, taken with permission from \citealt{AmaroSeoane18}), we see the conical velocity-space loss region at an instantaneous point along a star's elliptical orbit around a supermassive black hole.  The tidal radius (denoted $r_\mathrm{tid}$ in this image) will overlap with the star's pericentre if its instantaneous velocity vector $\vec{v}$ falls into the loss cone.  On the right (panel b, taken with permission from \citealt{Milosavljevic&Merritt03}), an orbit-averaged loss cone is depicted in the space of (negative) specific orbital energy $\mathcal E \equiv -E$ and angular momentum $j=\mathcal{R}^{1/2} = L/L_{\rm circ} \le 1$.  The loss cone becomes larger and larger in the space of dimensionless angular momentum as one moves to more tightly bound orbits (larger $|E|$).}
\label{fig:lossConeCartoons}
\end{figure}

The region of phase space with $L\le L_{\rm t}$ 
is colloquially called the loss cone (a term borrowed from plasma physics), which we illustrate in Figure~\ref{fig:lossConeCartoons}.
If angular momentum were conserved for every stellar orbit, the number of stars inside the loss cone would drop to zero within one orbital period. However, $L$ changes with time due to various processes: classical two-body and resonant relaxation, and torques induced by an aspherical stellar potential. Therefore, the time-averaged number of stars inside the loss cone is nonzero, and the tidal disruption rate per unit energy is
\begin{eqnarray}
\mathcal F_\mathrm{t}(E) &\equiv& \frac{1}{T_\mathrm{orb}(E)} 
\int_0^{\mathcal R_\mathrm{t}(E)} N(E, \mathcal R)\, \mathrm d \mathcal R, \\
T_\mathrm{orb}(E) &\equiv& T_\mathrm{orb}(E, \mathcal R=0), \qquad
\mathcal R_\mathrm{t} \equiv \left( \frac{L_\mathrm{t}}{L_\mathrm{circ}(E)} \right)^2 . \nonumber
\end{eqnarray}
Much of the remaining story revolves around estimating the equilibrium loss cone population, which is set by the efficiency of those processes that change the angular momenta of stars.
As a useful benchmark, consider a situation where these processes are extremely rapid, so that the angular momentum distribution is nearly uniform (isotropic). The corresponding rate of disruptions is called the (isotropic) full loss cone flux:
\begin{equation}  \label{eq:flux_full_iso}
\mathcal F_\mathrm{iso}(E) = \frac{\overline N(E)\, \mathcal R_\mathrm{t}(E)}{T_\mathrm{orb}(E)}.
\end{equation}
In a more complicated situation, the loss-cone population may be either smaller or larger than this reference value, depending on both the initial conditions and the efficiency of angular momentum mixing.

\subsection{Two-Body Relaxation}  \label{sec:relaxation}

A mechanism that operates in all stellar systems is two-body (or collisional) relaxation, caused by the discreteness of the stellar distribution. 
In the classical, ``Chandrasekhar,'' theory of two-body relaxation, the evolution of the stellar DF is driven by uncorrelated two-body encounters. 
We will begin, for simplicity, with a spherically symmetric nuclear star cluster surrounding a SMBH.  The star cluster has local three-dimensional density $n(r)$ and velocity dispersion $\sigma(r)$ in coordinate space (here $r$ is the distance from the SMBH), but its time evolution is most simply described by the Fokker--Planck equation in integral space, i.e. for $N(E, L, t)$.  We may define a local relaxation time as
\begin{equation}
    T_{\rm rel} \equiv \frac{\sigma^3}{G^2 n \langle m_\star^2 \rangle \ln\Lambda},
\end{equation}
where $\ln\Lambda \approx 10$ is the Coulomb logarithm \citep{Chandrasekhar42, SpitzerHart71}.  This expression assumes a present-day mass function (PDMF) of stars, ${\rm d}N/{\rm d}m_\star$, defined over the interval $[m_{\rm min}, m_{\rm max} ].$  The first and second moments of the PDMF are defined as
\begin{align}
    \langle m_\star \rangle =& \int_{m_{\rm min}}^{m_{\rm max}} m_\star \frac{{\rm d}N}{{\rm d}m_\star} {\rm d}m_\star  \notag   \\
    \langle m_\star^2 \rangle =&  \int_{m_{\rm min}}^{m_{\rm max}} m_\star^2 \frac{{\rm d}N}{{\rm d}m_\star} {\rm d}m_\star. \label{eq:PDMF}
\end{align}
Notably, the rate of two-body relaxation is controlled by the second moment, $\langle m_\star^2\rangle$, of the PDMF, indicating that for realistic stellar mass functions, it is typically the heaviest surviving species (often stellar-mass black holes) which dominate relaxation rates, especially at small radii where their relative fraction is higher due to mass segregation. For example, if one adds a population of stellar-mass black holes to a truncated Kroupa PDMF, the TDE rate can increase by up to a factor of $\approx 5$ \citep{StoneMetzger16}, depending on the assumed metallicity (which determines the relation between zero-age main sequence masses and compact remnant masses, e.g. \citealt{Belczynski+10}).

Over the characteristic relaxation timescale $T_\mathrm{rel} \gg T_\mathrm{orb}$, the typical change in energy is $\mathcal O\big(E\big)$ and in angular momentum is $\mathcal O\big(L_\mathrm{circ}(E)\big)$. However, far smaller changes in angular momentum $\mathcal O\big(L_\mathrm{t}\big)$ are needed to move stars on nearly radial orbits into or out of the loss cone: the time needed to produce characteristic changes in dimensionless angular momentum $\sim \mathcal{R}$ is $\sim \mathcal{R} T_{\rm rel}$. 
Because the relaxation of stars in angular momentum \textit{near the loss-cone boundary} occurs much faster than relaxation in energy \citep{Frank&Rees76}, in the first approximation these types of relaxation can be considered separately. As we will discuss shortly, this two-timescale decoupling establishes a (quasi-)steady-state distribution of stars in $L$ (or $\mathcal R$) at fixed $E$, which remains quasi-isotropic even in the presence of a loss cone \citep{Lightman&Shapiro77, Cohn&Kulsrud78}.

Over longer timescales, stellar diffusion in energy space gradually drives the stellar distribution towards a quasi-steady-state $\overline N(E)$ profile.  If the gravitational potential is dominated by the $\Phi \propto 1/r$ potential of the SMBH, this steady-state solution is known as the Bahcall--Wolf cusp, with  $f(E) \propto |E|^{1/4}$.  In coordinate space, the \citet{Bahcall&Wolf76} distribution translates to a spherically symmetric density profile\footnote{If a broad spectrum of stellar masses exist, it is generally the heaviest species that relaxes to the $n(r) \propto r^{-7/4}$ profile, while lighter species will achieve shallower, $n(r) \propto r^{-3/2}$ distributions \citep{Bahcall&Wolf77}. However, strong mass-segregation can give rise to steeper distributions, as in \citet{alexander&hopman09, Preto&AmaroSeoane10,Aharon&Perets2016}.} $n(r) \propto r^{-7/4}$, extending out to some fraction of the SMBH influence radius $r_{\rm inf}$, which we define to be the radius enclosing a total stellar mass equal to $M_\bullet$.  Notably, this is a {\it zero-flux solution} in energy space: in the limit as $R_{\rm t}/r_{\rm inf} \to 0$, the energy-space flux to the SMBH also goes to zero.

Accordingly, we will focus for now on the angular momentum diffusion\footnote{For a less rigorous -- but in some respects more intuitive -- approach operating entirely in coordinate space, see the work of \citet{SyerUlmer98}, which obtains qualitatively similar results to those presented here.}, by considering the one-dimensional, orbit-averaged Fokker--Planck equation for the evolution of the stellar distribution $N(\mathcal R,t)$ in the space of dimensionless angular momentum $\mathcal R$, at a fixed energy $E$ (the dependence of various quantities on $E$ is implied in the rest of this section, but is not explicitly marked):
\begin{equation}  \label{eq:diffusion_in_R}
\frac{\partial}{\partial t} N(\mathcal R,t) = \frac{\partial \mathcal F}{\partial \mathcal R}, \qquad
\mathcal F \equiv \mathscr D(E, \mathcal R)\,\frac{\partial N}{\partial \mathcal R}, \qquad
\mathscr D(E, \mathcal R) \approx \mathcal{D}(E)\, \mathcal R.
\end{equation}
As before, $\mathcal F$ is the flux of stars through integral space in the direction of the inner boundary\footnote{With this definition, the flux is positive if the stars diffuse towards the loss cone, as they usually do. Note that the sign convention is the opposite in some studies.} $\mathcal R_\mathrm{t}$, and the orbit-averaged diffusion coefficient $\mathcal D(E) \sim T_\mathrm{rel}^{-1}$ depends on various moments of the stellar DF \citep[Equation~6.31]{Merritt13}
\begin{align}  \label{eq:diffusion_coefficient}
\mathcal D(E) &= \frac{64\pi^2 \,G^2\; \ln\Lambda\; \langle m_\star^2\rangle}{3\, L_\mathrm{circ}^2(E)\,T_\mathrm{orb}(E)}\,
\int_0^{r_\mathrm{max}(E)} \!\!\!\!\!\mathrm{d} r\, \frac{r^2}{\sqrt{2[E-\Phi(r)]}} 
\big( 2 I_0 + 3 J_{1/2} - J_{3/2} \big) , \nonumber \\
I_0(E) &= \int_E^0 f(E')\, \mathrm{d} E' ,\\
J_n(E, r) &= \int_{\Phi(r)}^{E} \left[\frac{E'-\Phi(r)}{E-\Phi(r)} \right]^n f(E')\, \mathrm{d} E'.  \nonumber
\end{align}
Here $r_\mathrm{max}(E)$, the apocentre of a radial orbit with given specific energy, is the root of $\Phi(r_\mathrm{max})=E$. 
In a relatively shallow stellar density profile, or at high binding energies, $I_0(E)$ is the dominant moment of the stellar DF, and diffusion is a quasi-local process in energy space.  In a relatively steep stellar density profile, or at large radii, $J_n(E, r)$ will become of greater importance, and diffusion may become strongly non-local in energy space, with very tightly bound stars playing an important role in the evolution of more loosely bound bins of $E$. 
Over long timescales, the 1D Fokker--Planck approach of Eq.~(\ref{eq:diffusion_in_R}) may break down due to changes in the energy-space distribution of stars, $\bar{N}(E)$, and this breakdown will be hastened in bins of energy whose diffusion coefficients $\mathcal{D}(E)$ are dominated by contributions from distant regions of energy-space with shorter relaxation times.  In general, however, the 1D approach is a self-consistent way to determine quasi-steady state conditions near the loss cone boundary, since the time for orbits to experience a change in $\mathcal{R} \sim \mathcal{R}_{\rm t}$ is $\sim T_{\rm rel}(E)\mathcal{R}_{\rm t}$.

These equations are complemented with a no-flux, $\partial N / \partial \mathcal{R} = 0$ outer boundary condition (at $\mathcal R=1$), and an inner boundary condition (at $\mathcal R_\mathrm{t}$) of the Robin type (a linear combination of the function and its derivative):
\begin{equation}  \label{eq:boundary_condition}
N(\mathcal R_\mathrm{t}, t) - \alpha\, \mathcal R_\mathrm{t}\, 
\frac{\partial N}{\partial \mathcal R}\Big|_{\mathcal R=\mathcal R_\mathrm{t}} = 0,  \qquad
\alpha\approx (q^2+q^4)^{1/4},  \qquad
q\equiv \frac{\mathcal D\; T_\mathrm{orb}}{R_\mathrm{t}}.
\end{equation}
The dimensionless quantity $q(E)$ describes the diffusivity of relaxational loss cone refilling.
If the typical change of angular momentum per orbital time is larger than the size of the loss cone ($q\gg 1,\, \alpha\approx q$), the distribution of stars near and inside the loss cone is close to uniform (its gradient is small; $\partial N / \partial \mathcal R \ll N / \mathcal R$). In this case, we speak of a ``full loss cone'' or a ``pinhole'' regime where the flux through the loss-cone boundary is 
\begin{equation}  \label{eq:flux_full}
\mathcal F \approx \mathcal F_\mathrm{full} \equiv
\frac{N(\mathcal R_\mathrm{t}) \, \mathcal R_\mathrm{t}}{T_\mathrm{orb}}.
\end{equation}
This flux is proportional to the size of the loss cone but independent of the relaxation rate.
Note that, unlike Eq. (\ref{eq:flux_full_iso}) for the isotropic full loss cone flux, the above expression assumes only that the DF is \textit{locally} well-mixed near the loss-cone boundary.
In the opposite limit ($q\ll 1, \alpha \approx q^{1/2}$), we are in an ``empty loss cone'' or ``diffusive'' regime, where $N(\mathcal R_\mathrm{t})\approx 0$ and the flux is limited by the relaxation. 

We may consider a quasi-steady-state solution of Eq.~(\ref{eq:diffusion_in_R}) in which the shape of the DF $N(\mathcal R, t)$ stays the same, but its overall normalization $\overline N \equiv \int_0^1 N(\mathcal R)\,\mathrm{d}\mathcal R$ changes with time.  This solution has a nearly logarithmic profile \citep{Cohn&Kulsrud78}:
\begin{equation}  \label{eq:logarithmic_profile}
N(\mathcal R) \approx \overline N\, \frac
{\alpha + \ln(\mathcal R / \mathcal R_\mathrm{t})}
{(\alpha-1)(1-\mathcal R_\mathrm{t}) + \ln(1 / \mathcal R_\mathrm{t})},
\end{equation}
and the corresponding flux is
\begin{equation}  \label{eq:flux_general}
\mathcal F =
\frac{\mathcal D\;\overline N}{(\alpha-1)(1-\mathcal R_\mathrm{t}) + \ln(1 / \mathcal R_\mathrm{t})} =
\frac{q\; \mathcal F_\mathrm{full}}{(\alpha-1)(1-\mathcal R_\mathrm{t}) + \ln(1 / \mathcal R_\mathrm{t})}.
\end{equation}
As expected, in the full-loss cone limit the DF is nearly uniform ($N(\mathcal R) \approx \overline N$), and the flux approaches the isotropic full-loss-cone value (Eq. \ref{eq:flux_full_iso}), which coincides with Eq.~(\ref{eq:flux_full}) for a steady-state solution (however, this is no longer true for a time-dependent solution with strongly anisotropic initial conditions, as we will discuss in Section~\ref{sec:aniso}).
In the opposite, empty-loss cone limit, the DF falls sharply to zero for $\mathcal R \lesssim \mathcal R_\mathrm{t}$, and the flux is
\begin{equation}  \label{eq:flux_empty}
\mathcal F_\mathrm{empty} \equiv \frac{\mathcal D\;\overline N}{\ln(1 / \mathcal R_\mathrm{t}) + R_\mathrm{t} - 1}. 
\end{equation}
This empty loss cone flux is proportional to the relaxation rate but only weakly dependent on the size of the loss cone.  The steep drop in the DF below $R_\mathrm{t}$ implies that the TDE rate will be dominated by grazing encounters with $\beta \approx 1$ in the empty loss-cone regime (see Section~\ref{sec:synthetic}).

The transition between empty and full loss cone regimes corresponds to $\mathcal F \approx \frac{1}{2}\mathcal F_\mathrm{full}$, or $q \approx \ln(1/\mathcal R_\mathrm{t}) \sim \mathcal O(10)$. This occurs at the ``critical'' energy $E_{\rm crit}$, or the corresponding radius $r_{\rm crit}$.  The total TDE rate\footnote{In general, closed-form expressions for the variables in this section do not exist.  However, for the special case of a singular isothermal sphere density profile, with $n(r) \propto r^{-2}$, \citet{WangMerritt04} provide analytic expressions for $q(E)$, $\mathcal{F}(E)$, and $\dot{N}$.}
\begin{equation}
    \dot{N} \equiv \int_{-\infty}^0 \mathcal{F}(E){\rm d}E \label{eq:total_rate}
\end{equation}
will contain contributions from both empty- and full-loss cone regimes.  If $r_{\rm crit} \ll r_{\rm infl}$, then most of the total loss cone flux arises from $E \sim E_{\rm crit}$, and both the empty- and full-loss cone regimes contribute an $\mathcal{O}(1)$ fraction of the TDE rate\footnote{An interesting caveat to this discussion concerns extremely steep stellar cusps, i.e. those with density profiles falling off faster than $n(r) \propto r^{-9/4}$.  Such density profiles are not self-consistent, because they predict that, as $E\to -\infty$, $\mathcal{F}_{\rm empty}\to \infty$ \citep{SyerUlmer98}.  Density profiles of this steepness are rarely seen in nature, with the possible exception of post-starburst galaxies, which we discuss later \S \ref{sec:poststarburst}.}.  Conversely, if $r_{\rm crit} \gg r_{\rm infl}$, then most of the integrated loss cone flux comes from $E \sim -G M_\bullet / (2r_{\rm infl})$, and is predominantly in the empty loss cone regime \citep{SyerUlmer98}.  Astrophysical galactic nuclei typically possess $r_{\rm crit} \sim r_{\rm infl}$ \citep{WangMerritt04}, as we will discuss later in \S \ref{sec:empirical}.

\subsection{Anisotropic and Time-Dependent Loss Cones}
\label{sec:aniso}

So far, we have estimated the loss cone flux $\mathcal{F}(E)$ and TDE rate $\dot{N}$ in a relatively old galactic nucleus, one which has reached a quasi-steady state, quasi-isotropic distribution of orbital angular momentum in each bin of orbital energy (i.e. Eq. \ref{eq:logarithmic_profile}).  This quasi-steady state solution is attained on a timescale $\sim T_\mathrm{rel}$. At earlier times, the angular momentum distribution may be quite anisotropic, and in this regime, the capture rate depends sensitively on the initial conditions.  In certain types of galactic nuclei, where phenomena other than two-body relaxation are at work, it is not even clear that we can expect the angular momentum distribution to converge to Eq. \ref{eq:logarithmic_profile}.  For example, a galactic nucleus with a long-lived SMBH {\it binary} will preferentially eject stars on radial orbits, continually pumping up the tangential anisotropy of the star cluster \citep{Milosavljevic&Merritt03, Merritt&Wang05}.  The presence of a massive gas disc (i.e. an {\it active} galactic nucleus) will affect stellar orbits in a more complicated way \citep{Karas&Subr07}. The evolution of the nuclear stellar cusp and its build-up through star formation and/or cluster infall will also affect TDE rates \citep{Aharonetal2016}.  In this section, we will ignore these complications, and focus on DFs $f(E, \mathcal{R}, t)$ for spherical galactic nuclei evolving only due to two-body relaxation.  We consider more general loss-cone physics in aspherical potentials in \S \ref{sec:aspherical}.

In the limit of spherical symmetry and arbitrary initial conditions $f(E, \mathcal{R}, 0)$, it is possible to use Fourier-Bessel synthesis techniques to derive exact, time-dependent solutions to the Fokker--Planck equation in angular momentum space.  These solutions can then be converted into TDE rates via Eq. \ref{eq:diffusion_in_R} ($\mathcal{F} \propto \partial N / \partial \mathcal{R}$).  The semi-analytic Fourier-Bessel solutions were first derived in the empty loss cone limit \citep{Milosavljevic&Merritt03}, but can also be applied for more general inner boundary conditions \citep{Lezhnin&Vasiliev15}. 
While these solutions are useful, they are too lengthy to reproduce and explore in this review, so we will focus instead on results from the numerical solution of the time-dependent Fokker--Planck equation (Eq. \ref{eq:diffusion_in_R}).

The anisotropy of a stellar distribution can be quantified with the parameter (usually called $\beta$, but here we use $b$ to avoid confusion with Eq.~\ref{eq:beta})
\begin{equation}
    b(E) = 1-\frac{T_\perp(E)}{2T_\parallel(E)},
\end{equation}
which is a function of the total radial ($T_\parallel$) and tangential ($T_\perp$) kinetic energies of stellar orbits.
This definition can be related to the DF as $f(E, \mathcal{R}) = f_E(E)\mathcal{R}^{- b(E)}$, where $f_E(E)$ is a function independent of angular momentum, although more complicated (non-separable) DFs can have the same level of anisotropy.
When $b=0$, orbital velocities and angular momenta are distributed isotropically; when $b>0$, there is a radial orbit bias; when $b<0$, there is a tangential orbit bias.  In the case of a purely isotropic initial distribution ($b=0$; flat in $\mathcal{R}$), the initial TDE rates $\mathcal{F}(E) \approx \mathcal{F}_{\rm iso}(E)$ (Eq.~\ref{eq:flux_full_iso}), even at high binding energy where $q(E)\ll 1$.  In these energy bins, stars at $\mathcal R < \mathcal R_\mathrm{t}$ are removed on a timescale $T_\mathrm{orb}$, creating a steep gradient near the loss-cone boundary.  As the angular momentum distribution is progressively depleted at small but increasing $\mathcal R$ and approaches the Cohn--Kulsrud quasi-steady state profile (Eq.~\ref{eq:logarithmic_profile}), the gradient $(\partial N / \partial \mathcal R)_{\mathcal R = \mathcal R_\mathrm{t}}$ will soften, and the rates will decline and approach the steady-state value (Eq.~\ref{eq:flux_general}).  Because the Cohn--Kulsrud quasi-steady state is nearly isotropic (more specifically, a logarithmic function of $\mathcal{R}$), isotropic initial conditions produce time-dependent TDE rates that are not far from the steady-state expectation of Eq. \ref{eq:flux_general}.

Larger deviations from Eq.~\ref{eq:flux_general} will occur if the initial conditions are strongly anisotropic.  For example, a galactic nucleus may inherit a strong tangential anisotropy in the aftermath of a SMBH binary merger.  As two SMBHs inspiral, they eject those stars which pass within the orbit of the binary, scouring out a cavity in angular momentum space and depleting radial orbits \citep{Milosavljevic&Merritt03}.  This creates a gap in the initial distribution $N(\mathcal R, t=0)$ for $\mathcal R \le \mathcal R_\mathrm{gap} \gg \mathcal R_\mathrm{t}$.  After the SMBHs have merged, the TDE rate will be suppressed until the angular momentum gap is refilled, typically over a timescale $\sim 10^{-2} - 10^{-1}\,T_\mathrm{rel}$ \citep{Merritt&Wang05, Lezhnin&Vasiliev15}.

The opposite situation occurs when the initial distribution has an excess of stars with low angular momentum (a radially-biased velocity distribution).  This may occur naturally\footnote{But see also \citet{Arca&Capuzzo17} for counterexample simulations, where star cluster infall leads to a tangential bias.  Ultimately, the final $b(E)$ profile is likely sensitive to the orbital properties of infalling star clusters.} as a result of nuclear cluster buildup through infalling globular clusters (\citealt{Hartmann+11}, although many clusters are likely to disrupt far from the SMBH; \citealt{Perets&Mast2014}).  In the case of eccentric cluster infall, stars will be left behind on preferentially radial orbits.  
If we assume an idealized, initial radial anisotropy of $b$ across all bins of energy, then the TDE rate will initially be elevated and then decline with time roughly as $\mathcal F(E) \approx \mathcal F_\mathrm{full}(E) \times [ t / T_\mathrm{rel}(E) ]^{-b}$ \citep{Stone+18}. Despite the initially high rates of this scenario, this formula demonstrates (since $b \le 1$ definitionally) that the integrated total number of TDEs will be dominated by late times, once the distribution has become quasi-isotropic.

\subsection{Asphericity and Collisionless Loss Cone Refilling}
\label{sec:aspherical}

The classical loss-cone theory of Sections \ref{sec:relaxation} and \ref{sec:aniso} was developed in the 1970s for globular clusters, which are nearly spherical systems.  However, galactic nuclei are, to some extent, non-spherical (see e.g. \citealt{Lauer+05}), and stellar orbital angular momentum is, therefore, not conserved even in the absence of two-body relaxation.  Thus, even if the time-averaged angular momentum $\overline L$ of a star on a given orbit is large, the minimum value of angular momentum $L_\mathrm{min}$ reachable by this orbit may be smaller than $L_\mathrm{t}$, potentially bringing a much larger reservoir of stars into the loss cone.

In perfectly axisymmetric systems, one component of angular momentum ($L_z$) is still conserved, so $L_\mathrm{min} \ge L_z$; however, in triaxial or even less symmetric potentials, $L_\mathrm{min}$ may be zero for a significant fraction of orbits; specifically, those in ``centrophilic'' orbit families, such as box, pyramid or chaotic orbits \citep{Poon&Merritt01}.  More generally, we refer to all orbits with $L_\mathrm{min} \le L_\mathrm{t}$ as the ``drain region\footnote{The drain region is often called the ``loss wedge'' in the axisymmetric case -- e.g. \citet{Magorrian&Tremaine99}.},'' and denote the fraction of phase space occupied by these orbits at a given energy as $\mathcal R_\mathrm{drain}(E)$. Most of these orbits have low average angular momentum, i.e. $\overline L \ll L_\mathrm{circ}(E)$, although not all orbits with $\overline L \ll L_\mathrm{circ}(E)$ are centrophilic. Nevertheless, for the sake of simplicity, we assume that the drain region is simply $\mathcal R < \mathcal R_\mathrm{drain}$.

There are two separate effects associated with the existence of a drain region.
First, the scalar angular momentum changes significantly (by $\mathcal O(\epsilon\,L_\mathrm{circ}))$ on a timescale $T_\mathrm{ang}\simeq \epsilon^{-1}\,T_\mathrm{prec}$, where $\epsilon$ is the (dimensionless) degree of non-sphericity (the relative difference between the three principal axes). Here the period of pericentre precession $T_\mathrm{prec}\equiv 2\pi(\Omega_r-\Omega_\phi)^{-1}$ is given by the difference between the frequencies of the radial and azimuthal motion: it is comparable to the orbital period $T_\mathrm{orb}\equiv 2\pi\Omega_r^{-1}$ outside the SMBH radius of influence, but is much longer than $T_\mathrm{orb}$ in a nearly Keplerian potential (see Equation~4.88 in \citealt{Merritt13}).
In the drain region, the timescale for the scalar angular momentum to change by $\mathcal O(L_\mathrm{t})$ is typically shorter than $T_\mathrm{orb}$. Therefore, stars on drain orbits are in the full loss cone regime \citep{Merritt&Poon04}, and their capture rate is given by Eq.~(\ref{eq:flux_full}). Note that this does not imply that the capture rate is equal to the isotropic full loss cone rate of Eq.~(\ref{eq:flux_full_iso}), because the density of stars in the drain region $N(E, \mathcal R < \mathcal R_\mathrm{drain})$ may be different from the angular momentum-averaged density of stars $\overline N(E)$.
We may now define the ``drain timescale'' needed to remove the stars from this region: $T_\mathrm{drain} \equiv T_\mathrm{orb}\, \mathcal R_\mathrm{drain} / \mathcal R_\mathrm{t}$.
It turns out that in axisymmetric systems, this time is still much shorter than the Hubble time, at least in those galactic nuclei with $M_\bullet \lesssim 10^8\,M_\odot$ that are the main sources of TDEs. However, in triaxial systems, the fraction of centrophilic orbits, $\mathcal R_\mathrm{drain}$, could be $\gtrsim 0.1$ (it is proportional to the degree of non-sphericity $\epsilon$), making the draining time longer than the Hubble time for most orbits (see Equations~6,~7 and Figure~4 in \citealt{Vasiliev14}). During this time, the capture rate is determined by the initial conditions.  In the ``most neutral'' case of a (nearly-)isotropic distribution in angular momentum, the TDE rate is of order the isotropic full loss cone rate of Eq.~(\ref{eq:flux_full_iso}), which could be much higher than the relaxation-limited capture rate in purely spherical nuclei \citep{Magorrian&Tremaine99}.

\begin{figure}
\includegraphics{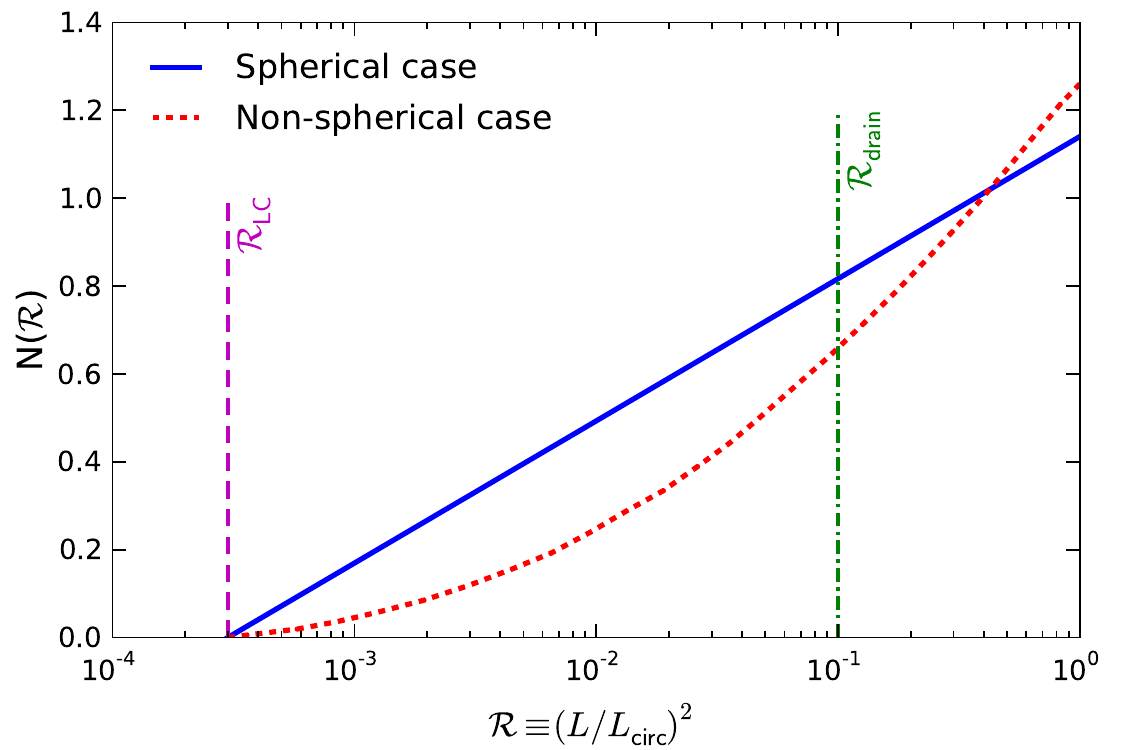}
\caption{
Illustration of the quasi-steady-state DF in spherical (solid blue) and non-spherical (dotted red) galactic nuclei, for the case of an empty loss cone ($q=0$ in Eq. ~\ref{eq:boundary_condition}) with a fairly large loss-cone size ($\mathcal R_\mathrm{t}=3\times10^{-4}$). In the spherical case, the DF has a logarithmic profile (Eq. \ref{eq:logarithmic_profile}), and the slope of this profile in the semi-log-scaled coordinates determines the flux. In the non-spherical case, the profile flattens out at $\mathcal R \lesssim \mathcal R_\mathrm{drain}$ (${}=0.1$ in this example), because the angular momenta of stars are shuffled by non-spherical torques. At the same time, the profile at $\mathcal R \gtrsim \mathcal R_\mathrm{drain}$ remains nearly logarithmic, as it is still determined by relaxation-driven diffusion, but it has a steeper slope (roughly a factor of two in this example), producing a correspondingly larger flux. Adapted from Figure~5 of \citealt{Vasiliev&Merritt13}.
}
\label{fig:losscone_nonsph}
\end{figure}

After the initial population of stars in the drain region has been depleted, two-body relaxation again becomes the rate-limiting step. However, relaxation will still be aided by the existence of the drain region, which acts as an ``extended'' loss cone: one with a much larger surface area in phase space.  After a star has diffused into the drain region via collisional relaxation, it will collisionlessly wander into the actual loss cone in a time $\lesssim T_\mathrm{drain}$ (unless it is scattered back into the higher-$\overline L$ region of phase space).
The TDE rate from the loss wedge is technically still set by the full loss-cone rate of Eq.~(\ref{eq:flux_full}), but the phase-space density of stars in this region, $N(\mathcal R_\mathrm{t}) \simeq N(\mathcal R_\mathrm{drain})$, may be much lower than $\overline N$, if the supply rate is diffusion-limited. In this case, the steady-state flux is given by the expression for the empty-loss cone regime of Eq.~(\ref{eq:flux_empty}), but replacing $\ln(1/\mathcal R_\mathrm{t})$ with $\ln(1/\mathcal R_\mathrm{drain})$. Because of this logarithmic dependence, the actual enhancement of capture rate is at most a factor of few, as is illustrated by Figure~\ref{fig:losscone_nonsph}.

A different type of transient asphericity
can occur in special, ``degenerate'' potentials where frequencies of motion are rationally commensurate with each other and orbits close.  For example, in the Kepler potential of the SMBH, all three frequencies of motion are exactly equal for an individual star (so long as we neglect precession from the background stellar potential and relativistic effects).  Because the stellar cusp is made of a finite number of stars, the combination of discreteness and closed orbits will create a statistical excess of time-averaged stellar mass in one direction: a temporary asphericity.  An alternative view of this process is that pairwise encounters between nearby stars will be {\it correlated} over time, and therefore may coherently torque stellar orbits much more efficiently than do the {\it uncorrelated} effects from two-body relaxation.  The resulting orbital evolution, known as {\it resonant relaxation} \citep{Rauch&Tremaine96}, has been proposed as a way to more efficiently refill empty loss cones \citep{Hopman&Alexander06}.  However, general relativistic precession often prevents resonant relaxation from exciting stars to the high eccentricities needed to enter the loss cone \citep{Merritt+11, Brem+14}, and recent calculations suggest that its impact on the loss-cone flux is fairly minor in realistic galactic nuclei (e.g., \citealt{Merritt15}, or Fig. 17 in \citealt{BarOr&Alexander16}).

To summarize, the TDE rate in non-spherical systems is larger than in spherical ones, if most of the flux in the latter is delivered in the diffusion-limited (empty loss cone) regime.  In Section \ref{sec:empirical}, we will see that in denser galactic nuclei, with $M_\bullet \lesssim 10^6\,M_\odot$, this enhancement is small, while for $M_\bullet\gtrsim 10^8\,M_\odot$ it could be more than an order of magnitude.  However, these larger galaxies are (i) rarer and (ii) have difficulty producing luminous TDEs ($M_\bullet \sim M_{\rm H}$), and so do not dominate the cosmic TDE rate of main-sequence stars. 

\subsection{General Relativistic Loss Cone Theory}
\label{sec:GR}

Einstein's theory of general relativity (GR) differs from Newtonian gravity in several respects that may carry important observational consequences for TDEs.  The first of these differences concerns the nature of gravity itself in the two theories.  In Newtonian
gravity, the SMBH exerts an inverse-square-law force on the star, pulling more strongly on the side of the star facing the SMBH than on the stellar centre of mass.  Equating the difference in the acceleration experienced by the stellar centre of mass and surface (i.e. the tidal acceleration) to the star's self gravity and solving for the distance from the SMBH yields the Newtonian tidal radius $R_{\rm t}$ (Eq.~\ref{eq:rtidal}).  

General relativistic gravity is  interpreted instead as a non-vanishing spacetime curvature that determines the geodesics along which test particles travel.  This spacetime curvature can cause initially parallel geodesics to deviate, leading to tidal disruption if the rate of geodesic deviation exceeds the star's self-gravity.  The relativistic geodesic-deviation equation is most conveniently expressed in Fermi normal coordinates $\{ \tau, X^{(i)}\}$ \citep{Manasse&Misner+63,Marck+83,1983grg1.conf..438L}.  Here $\tau$ is the proper time along the central timelike geodesic on which the star's centre of mass travels, and $X^{(i)}$ are Cartesian spatial coordinates in a spatial hypersurface orthogonal to this central geodesic.  In these coordinates, the geodesic-deviation equation becomes
\begin{equation} \label{E:geoDFNC} 
\frac{d^2X^{(i)}}{d\tau^2} = -C^{(i)}_{\quad(j)} X^{(j)}.
\end{equation}
where 
\begin{equation} \label{E:TidalTens}
C^{(i)}_{\quad(j)} = R^\beta_{~\mu\alpha\nu} \lambda_\beta^{~(i)} \lambda^\mu_{~(0)} \lambda^\alpha_{~(j)} \lambda^\nu_{~(0)}
\end{equation}
is the tidal tensor, $R^\beta_{~\mu\alpha\nu}$ is the Riemann curvature tensor, and $\{ \lambda^\mu_{~(0)}, \lambda^\mu_{~(i)}\}$ is the orthonormal tetrad of 4-vectors with respect to which the Fermi normal coordinates $\{ \tau, X^{(i)}\}$ are defined.  The symmetries of the Riemann tensor imply that the tidal tensor is a real, symmetric $3\times 3$ matrix with three real eigenvalues and orthogonal eigenvectors.  Eq.~\ref{E:geoDFNC} implies that the star will be stretched along the eigenvector corresponding to the tidal tensor's sole negative eigenvalue, $V_-$ ; equating the tidal acceleration in this direction at pericentre to the star's self-gravity yields the relativistic generalization of the tidal radius \citep{Kesden12}
\begin{equation} \label{E:GRrtidal}
R_{\rm t,GR} =  \left[ \left( \frac{|V_-|}{2GM_\bullet/r^3} \right) \left( \frac{M_\bullet}{M_\star} \right) \right]^{1/3} R_\star.
\end{equation}
As in the Newtonian case (Eq. \ref{eq:rtidal}), the exact criterion for tidal disruption is $R_{\rm p} < R_{\rm t,GR}/\beta_{\rm crit}$, where $\beta_{\rm crit} \approx 1 - 2$ depends on the internal structure of the star. In the non-relativistic limit ($v \ll c, r \gg R_{\rm g}$), $R_{\rm t,GR}$ reduces to $R_{\rm t}$.  

Although there is a superficial similarity between the Newtonian and relativistic tidal radii of Eqs.~\ref{eq:rtidal} and \ref{E:GRrtidal}, there are also important differences.  Because the tidal tensor $C^{(i)}_{(j)}$ and thus its negative eigenvalue $V_-$ depend on both the Riemann tensor and the stellar geodesic, the relativistic tidal radius depends on both the spacetime metric of the SMBH and the stellar 4-position and 4-velocity at pericentre, not just the distance from the SMBH.  The spacetime near SMBHs is described in GR by the Kerr metric \citep{1963PhRvL..11..237K}, which depends not just on the SMBH mass $M_\bullet$, but also on its dimensionless spin, $0 \leq \chi_\bullet \leq 1$ \citep{1971PhRvL..26..331C}.  

The spin dependence of the SMBH's gravity is the second important difference between Newtonian gravity and GR.  The SMBH spin breaks the spherical symmetry present in Newtonian point potentials, implying that, in the commonly used Boyer-Lindquist coordinate system \citep{1967JMP.....8..265B}, the Riemann tensor depends on both the radial coordinate $r$ and the polar coordinate $\theta$.  The Kerr metric is stationary and axisymmetric, implying the existence of a specific energy $E$ and angular momentum $L_z$ that are conserved along geodesics.  In addition, the Kerr metric has a Killing tensor that provides a conserved Carter constant $Q$ \citep{1968PhRv..174.1559C,1970CMaPh..18..265W}.  In the non-relativistic limit, the Carter constant corresponds to the square of the magnitude of the component of the orbital angular momentum in the equatorial plane of the SMBH.  This allows us to define an (effective) specific orbital angular momentum $L \equiv \sqrt{L_z^2 + Q}$, and an (effective) inclination $\iota = \cos^{-1}(L_z/\sqrt{L_z^2 + Q})$ that are conserved along all Kerr geodesics.  These considerations indicate that the relativistic tidal radius $R_{\rm t,GR}$ can, in principle, depend on the SMBH mass $M_\bullet$ and spin $\chi_\bullet$, but also on the stellar orbital energy $E$, inclination $\iota$, and argument of pericentre $\omega$.  In practice, the dependence on $E$ and $\omega$ can be neglected to high accuracy, allowing us to define the threshold for tidal disruption $L_{\rm t,GR}(\chi_\bullet, \iota)$ as the value of $L$ for which
\begin{equation} \label{E:LtGR}
V_-(M_\bullet, \chi_\bullet, L, \iota) = -\frac{2GM_\star}{R_\star^3} 
\end{equation}
when evaluated at pericentre\footnote{The right-hand side of Eq.~\ref{E:LtGR} should be multiplied by $\beta_{\rm crit}^3$ to account for stellar structure.}.

A third significant difference between Newtonian gravity and GR is that, in the latter theory, a black hole is defined as an object possessing an event horizon, a hypersurface from within which even light cannot escape \citep{1984ucp..book.....W}.  The tidal force exerted by a Newtonian point mass scales $\propto r^{-3}$ and can therefore become arbitrarily large.  This implies that any Newtonian point mass is capable of tidal disruption, given a small enough pericentre.  However, to produce an observable TDE in GR, a SMBH's tides must overcome the star's self gravity while avoiding direct capture of the tidal debris by the event horizon.  Stars on parabolic orbits (and their resulting debris) will be captured when their specific orbital angular momentum $L$ falls below a threshold $L_{\rm cap}(\chi_\bullet, \iota)$ that depends on both the SMBH spin and orbital inclination.  The rate of observable TDEs in GR will therefore be given by the rate at which two-body relaxation or other processes described above drive stars onto orbits with $L_{\rm cap}(\chi_\bullet, \iota) < L < L_{\rm t,GR}(\chi_\bullet, \iota)$.  The hierarchy of distance scales, $R_{\rm g} \ll r_{\rm infl}$, between the gravitational radius and the radius of influence (from which most tidally disrupted stars are scattered into the loss cone) ensures that GR only modifies the boundaries of the loss cone, but does not otherwise affect the process responsible for refilling it.

The capture threshold $L_{\rm cap}(\chi_\bullet, \iota)$ scales linearly with $M_\bullet$ on dimensional grounds, while to lowest order in $M_\star/M_\bullet \ll 1$, the threshold for tidal disruption $L_{\rm t,GR}(\chi_\bullet, \iota) \propto M_\bullet^{2/3}$, like its Newtonian counterpart $\sqrt{2GM_\bullet R_{\rm t}}$.  This implies that SMBHs more massive than the relativistic Hills mass $M_{\rm H,GR}(\chi_\bullet)$, defined such that $L_{\rm cap}(\chi_\bullet, \iota) > L_{\rm t,GR}(\chi_\bullet, \iota)$ for all inclinations $\iota$, will be incapable of producing observable TDEs.  For Schwarzschild SMBHs ($\chi_\bullet = 0$), spherical symmetry is restored and the disruption and capture thresholds are inclination-independent.
The relativistic Hills mass for a Schwarzschild SMBH is
\begin{equation} \label{E:relHills}
M_{\rm H,GR} = \left( \frac{5c^6R_\star^3}{128\beta_{\rm crit}^3G^3M_\star} \right)^{1/2} = \left( \frac{5}{16\beta_{\rm crit}^3} \right)^{1/2} M_{\rm H} = 10^{7.8}\beta_{\rm crit}^{-3/2}M_\odot,
\end{equation}
where $\beta_{\rm crit}$ is the minimum penetration factor for tidal disruption and $M_{\rm H}$ is the Newtonian Hills mass given by Eq.~(\ref{eq:MHills}).  For high-mass stars like our Sun whose equation of state can be approximated by a polytropic index $\gamma = 4/3$, Newtonian hydrodynamic simulations suggest that $\beta_{\rm crit} \simeq 1.85$ \citep{Guillochon&RamirezRuiz13} implying $M_{\rm H,GR} \simeq 10^{7.4} M_\odot$ \citep{2017PhRvD..95h3001S}.
For Kerr SMBHs ($\chi_\bullet > 0$), both $L_{\rm t,GR}(\chi_\bullet, \iota)$ and $L_{\rm cap}(\chi_\bullet, \iota)$ are monotonically increasing functions of inclination $\iota$, but the direct capture threshold has a steeper dependence.  This implies that the Hills mass $M_{\rm H,GR}(\chi_\bullet)$ will be determined by the condition $L_{\rm cap}(\chi_\bullet, \iota = 0^\circ) = L_{\rm t,GR}(\chi_\bullet, \iota = 0^\circ)$.  For maximally spinning SMBHs ($\chi_\bullet = 1$), this limit can be as large as $\sim 10^9 M_\odot$ \citep{1994MmSAI..65.1135S,2006A&A...448..843I,Kesden12}. 

\begin{figure}
\includegraphics[width=120mm]{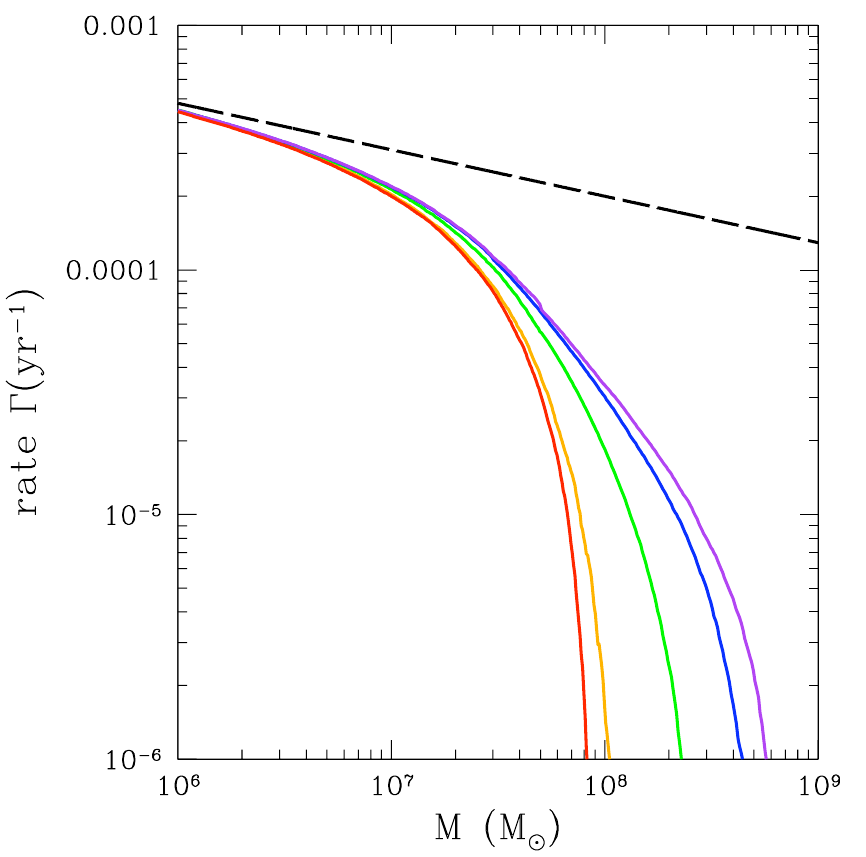}
\caption{The TDE rate $\Gamma$ in an idealized singular isothermal sphere ($\rho \propto r^{-2}$) galactic nucleus, shown as a function of SMBH mass $M$.  The black dashed line shows the analytic estimate from \citet{WangMerritt04} which neglects direct capture of stars by the event horizon.  Colored lines show TDE rate curves corrected for this general relativistic effect, with the red, orange, green, blue and purple curves corresponding to SMBH spin magnitudes $\chi_\bullet$ with values of $0.0$, $0.5$, $0.9$, $0.99$, and $0.999$, respectively.  These calculations average over an isotropic distribution of incoming stellar orbits and assume all disrupted stars are of solar mass and radius with $\beta_{\rm crit} = 2^{-1/3}$.  Taken with permission from \citet{Kesden12}.}
\label{fig:spinSuppression}
\end{figure}

We show the reduction in the observable TDE rate due to direct capture by the event horizon in Fig.~\ref{fig:spinSuppression} \citep{Kesden12}.  The dashed black curve shows the TDE rate as a function of SMBH mass $M_\bullet$ for galaxies with a singular isothermal sphere ($\rho \propto r^{-2}$) stellar density profile and velocity dispersions $\sigma$ set by the $M_\bullet-\sigma$ relation \citep{2000ApJ...539L...9F,2000ApJ...539L..13G}.  These rates were calculated with a Newtonian loss cone refilled by two-body relaxation as described in Sec.~\ref{sec:relaxation} 
\citep{WangMerritt04}.  The solid colored curves (corresponding to different SMBH spins as indicated in the caption) show the suppression of this TDE rate from the fraction of stars in the tidal disruption loss cone, $L < L_{\rm t,GR}(\chi_\bullet, \iota)$, that also lie within the direct capture loss cone\footnote{This calculation assumes the full loss-cone limit, in computing the relativistic correction factor, although the per-galaxy TDE rates shown in Fig. 3 are based on loss-cone calculations using contributions from both the full and empty regions.}, $L < L_{\rm cap}(\chi_\bullet, \iota)$.  We see that the event horizon has a negligible effect for $M_\bullet \lesssim 10^6 M_\odot$, and that SMBHs with $M_\bullet \approx 5\times 10^{8} M_\odot$ can still produce observable TDEs provided their spins are large enough.  The predicted suppression of the TDE rate by direct capture for $M_\bullet \gtrsim 10^{7.5} M_\odot$ shown in Fig.~\ref{fig:spinSuppression} is consistent with the super-exponential cutoff in the TDE rate observed in a limited sample of twelve optically-selected TDE candidates \citep{vanVelzen18}.

Calculating observable TDE rates in a fully self-consistent manner with the asymmetric, relativistic loss cones described above remains an open problem, but we can anticipate several qualitative features of the results.  For Schwarzschild SMBHs, the tidal acceleration is stronger in GR than on Newtonian orbits with the same angular momentum $L$ \citep{2017PhRvD..95h3001S}.  Eq.~\ref{eq:flux_full} thus implies that, in the full loss-cone regime, the TDE rate will be enhanced by a factor of $(L_{\rm t,GR}/L_{\rm t})^2$ in the absence of direct capture by the event horizon.  This factor is a monotonically increasing function of SMBH mass that reaches $\approx 2.7$ for $M_\bullet = M_{\rm H,GR}(\chi_\bullet = 0)$, though by definition direct capture cannot be neglected for SMBHs near the Hills mass.  
In the {\it empty} loss cone regime, direct capture can be generally neglected so long as $L_{\rm t, GR}(\chi_\bullet, \iota) > L_{\rm cap, GR}(\chi_\bullet, \iota)$, but in this case the TDE rate enhancement is only by a factor $\approx \ln(L_{\rm t,GR}/L_{\rm t})$, as in Eq. \ref{eq:flux_empty}.  However, according to Eq. \ref{eq:boundary_condition}, the dimensionless factor $q(E)$ will be suppressed in GR by the same factor $(L_{\rm t,GR}/L_{\rm t})^2$, pushing more of the phase space into the empty loss-cone regime where the TDE rate is suppressed with respect to the full loss-cone regime by this factor.  In either loss cone regime, these effects will be small when $M_\bullet \ll M_{\rm H,GR}$.

SMBH spin further complicates TDE rate predictions.  The thresholds $L_{\rm t,GR}(\chi_\bullet, \iota)$ and $L_{\rm cap}(\chi_\bullet, \iota)$ are smaller for Kerr SMBHs than their Schwarzschild values when orbits are prograde ($\iota < 90^\circ$), and are larger when orbits are retrograde ($\iota > 90^\circ$).  For isotropic distributions (flat in $\cos\iota$), even large spins have a modest $\lesssim 10\%$ effect on the $M_\bullet$-integrated TDE and capture rates in the full loss-cone regime \citep{Young+77}.  However, we have already seen that spin can have {\it dramatic} effects on rates of observable TDEs from individual bins of SMBH mass when $M_\bullet \sim 10^8 M_\odot$ (Fig. \ref{fig:spinSuppression}).  Furthermore, spin can have a significant effect on the inclination distributions of tidally disrupted and captured stars, with important observational consequences that we discuss further in Section \ref{sec:synthetic}, and in greater detail in the \flowchap{}. 

\subsection{Other Loss Cones}
\label{sec:types}
So far, we have focused our attention on the loss cone relevant for standard TDEs: one centered on a massive black hole, with a boundary defined by the complete tidal disruption of a main sequence star.  But the concept of a loss cone can be applied more generally to compute rates of other types of tidal disruptions.  Some, such as the disruptions of binary or giant-branch stars by SMBHs, have event rates set by loss cone considerations not too dissimilar from those discussed already.  More exotic types of tidal disruptions, such as ``micro-TDEs'' involving hyperbolic flybys of stars and stellar-mass black holes \citep{Peretsetal2016}, or the short gamma-ray bursts produced by the quasi-circular inspiral of a neutron star into a stellar-mass black hole, are produced by quite different dynamical processes, and are therefore outside the purview of this Chapter.  In this section, we briefly overview how loss cone physics is altered for different types of tidal disruptions: partial rather than full (\S \ref{sec:partials}), giant-branch rather than main sequence (\S \ref{sec:giants}), and binary rather than single (\S \ref{sec:binaries}).  Stars may also be tidally disrupted by {\it binary} SMBHs, but the underlying dynamics here are sufficiently complicated that they are left to the \binchap.

\subsubsection{Partial Disruptions}
\label{sec:partials}
Partial disruptions will occur when stars approach the central SMBH with $\beta < \beta_{\rm crit}$.  As discussed previously, the exact value of $\beta_{\rm crit}$ is a number $\mathcal{O}(1)$ that depends on the internal structure of the victim star.  For simple polytropic models, $\beta_{\rm crit}$ ranges from $\approx 0.92$ (for a relatively fluffy, $n=3/2$ polytrope, representative of lower main sequence stars) to $2.01$ (for a more centrally concentrated $n=3$ polytrope, representative of Sun-like stars).  While these thresholds for full disruption are well-established for polytropic stellar models \citep{Guillochon&RamirezRuiz13, Mainetti+17}, the corresponding $\beta_{\rm crit}$ values for more realistic stellar structures have yet to be determined.  The exact threshold below which {\it no} mass loss occurs is also a function of stellar structure; for polytropic models, mass loss typically requires $\beta > \beta_{\rm min} \approx 0.5$ \citep{Guillochon&RamirezRuiz13}.  

Because the cross-section for partial disruption is substantially larger than that for full disruption, partial disruptions should be more common.  This is clearly true in the empty loss-cone limit; when $q(E) \ll 1$, higher $\beta$ values are exponentially suppressed\footnote{Although we note that a power-law tail of high-$\beta$ TDEs will occur, even when $q \ll 1$, due to the effects of strong scattering \citep{Weissbein&Sari17}.}.  In the full loss-cone regime, the number of stars $N(\mathcal{R})$ is roughly independent of $\mathcal{R}$ deep into the loss cone, meaning that the differential rate of disruptions ${\rm d}\dot{N}/ {\rm d}\mathcal{R} \propto {\rm d}\dot{N}/ {\rm d}r_{\rm p} \propto \mathrm{const}$.  By a change of variables, this gives the differential rate ${\rm d}\dot{N}/ {\rm d}\beta \propto \beta^{-2}$.  The ratio of partial to full tidal disruptions will, in the $q(E)\gg 1$ limit, be
\begin{equation}  \label{eq:Ndot_partial}
\frac{\dot{N}_{\rm partial}}{\dot{N}_{\rm full}} = \frac{\beta_{\rm min}^{-1} - \beta_{\rm crit}^{-1}}{\beta_{\rm crit}^{-1} - \beta_{\rm max}^{-1}},
\end{equation}
where $\beta_{\rm max}$ is the maximum penetration parameter that can avoid direct capture by the horizon.  In Newtonian gravity, where the horizon may be approximated as an absorbing boundary at $2R_{\rm g}$, 
\begin{equation}
\beta_{\rm max}^{\rm N} = \frac{R_\star c^2}{2GM_\bullet^{2/3}M_\star^{1/3}}. 
\end{equation}
In GR, $\beta_{\rm max} = \beta_{\rm max}^{\rm GR}(\chi_\bullet, \iota)$, and can be computed by determining both $L_{\rm t, GR}$ and $L_{\rm cap}$, although it is important to note that the definition of $\beta$ becomes more ambiguous in relativistic gravity \citep{2017PhRvD..95h3001S}.  For simplicity, we have so far discussed differential rates ${\rm d}\dot{N}/{\rm d}\beta$ in the two extreme limits of loss cone repopulation; a more sophisticated treatment of the intermediate, $q \sim 1$ case can be found in \citet{Strubbe11}.

\subsubsection{Giant Stars}
\label{sec:giants}
After a sufficient fraction of their hydrogen has been burnt, most stars will evolve off the main sequence and become giants, increasing their radial size by at least one order of magnitude.  Stars with initial mass $M_\star \sim 1-8 M_\odot$ will spend most of their post-main sequence evolution on the red giant branch (RGB), with $R_{\rm RG} \sim 10R_\odot$, and a smaller but dynamically important portion on the AGB branch with $R_{\rm AGB} \sim 100R_\odot$.  

Giant-branch stars are, consequently, much more vulnerable to partial tidal disruption: while their cores are no less dense than those of main sequence stars, their envelopes are distended and only weakly bound.  If we reapply the results of \S \ref{sec:relaxation}, we therefore expect that per-star TDE rates of giants should be larger by a factor $\sim R_{\rm G} / R_\star$ in the full loss cone regime, but by a more modest factor $\sim \ln(R_{\rm G} / R_\star)$ in the empty loss cone regime.  Because the total TDE rate $\dot{N}$ is an integral across these two regimes, it will typically follow a sublinear power law as shallow as $\dot{N} \propto (R_{\rm G} / R_\star)^{1/4}$ \citep{MacLeod+12}.  While these arguments show that the per-star rate of giant disruption is higher than that for main sequence stars, this enhancement competes unsuccessfully with the much smaller number of giant-branch stars.  As a result, the ratio between giant-branch and main sequence TDE rates is $\dot{N}_{\rm G} / \dot{N}_{\rm MS} \sim 0.1$ \citep{Magorrian&Tremaine99, MacLeod+12}.  This ratio holds across a wide range of SMBH masses but breaks down above the main sequence Hills mass, when $M_\bullet \gtrsim M_{\rm H}$.  Schwarzschild SMBHs with $10^8 M_\odot \lesssim M_\bullet \lesssim 10^9 M_\odot$ are generally unable to disrupt main sequence stars, and only produce luminous flares from giant disruptions.  When $M_\bullet \gtrsim 10^9 M_\odot$, Schwarzschild SMBHs will become incapable of disrupting most giant-branch stars as well, and their only luminous flares will come from the small population of stars at the tips of the RGB and AGB \citep{MacLeod+12}.

The time-dependent radii of giants gives these stars a non-diffusive way to enter the loss cone, one that is inaccessible to main sequence stars: expanding in size until their loss cone grows to intersect their current orbit.  ``Growth into the loss cone'' was first investigated by \citet{SyerUlmer98}, who argued that this will be the dominant source of TDEs from evolved stars.  However, this early work assumed that every giant disruption would be a full and highly luminous one.  In reality, stars that grow onto the loss cone will, very likely, be ``spoon-fed'' to the SMBH in a series of tens to hundreds of very weak partial disruptions \citep{MacLeod+13}, making these events quite challenging to detect.

We note, however, that various dynamical processes can shorten relaxation times outside the influence radius \citep{Perets+07a,Hamers:2017}. At these regions MS stars are typically already in the full loss cone regime and are not significantly affected by such processes. However, the empty loss cone for objects with larger tidal radii such as binary stars and giant stars extend to larger distances (as we discuss in the next section). Such processes can therefore increase the TDE rates of giants, and change the above-mentioned ratio, leading to a greater contribution of TDEs from giant stars.  

\subsubsection{Binary Stars}
\label{sec:binaries}
Binary stars -- especially massive ones -- can be as common as single stars. Their fate, when plunging along highly eccentric orbits towards a SMBH, can be richer than that of single stars. A binary star passing near a SMBH can be tidally separated with no tidal disruption of the single-star components. In this case, the individual stars may either recombine to reform a binary after pericentre passage, or undergo a three-body exchange interaction, with one star being captured around the black hole and the other being ejected with velocity in excess of the bulge escape speed \citep{hills88,hills91,SKR10}. These escapers may have already been observed as hypervelocity stars (HVSs) in the halo of our Galaxy \citep[e.g.][]{brownw18}. Such outcomes most likely occur for binaries with internal semimajor axes $a_{\rm bin}\gg R_\star$, with a centre-of-mass orbit just grazing the tidal separation radius $R_{\rm sep} = a (M_\bullet/m _{\rm bin})^{1/3}$ (here $m_{\rm bin}$ is the binary mass). Tidal disruption, or tidally-induced mergers, may result from either deeper encounters that cross the stellar tidal radius, $R_{\rm t} \approx (R_\star/a_{\rm bin}) R_{\rm sep}$, or near-contact binaries for which $a_{\rm bin} \sim R_\star$ \citep{mandel&levin15,bradnick+17}. For details on how the disruption and accretion processes are altered by binarity, we refer the reader to the \disrupchap{} and the \flowchap{}. Here we will discuss properties of the stellar binaries' loss cone, the ensuing event rates, and their connections with HVSs.

As discussed in \S\ref{sec:relaxation}, the critical radius measures the size of the region that is in the empty loss cone regime, and it depends on the tidal radius of the system under consideration (and, strictly speaking, can be more precisely defined in energy space, a detail we neglect in this section).  When the critical radius is smaller than the sphere of influence radius, $r_{\rm crit} \propto R_{\rm t}$, but in the opposite case, when $r_{\rm crit} > r_{\rm inf}$, $r_{\rm crit} \propto R_{\rm t}^{1/(4-\gamma)}$, where $\gamma$ is the power-law slope of the nuclear stellar density profile \cite[e.g. Eqs. 9-10 in][]{SyerUlmer98}. Empirically, observed nuclear density profiles typically have $0\lesssim \gamma \lesssim 2$ \citep{Lauer+05}; in the Milky Way specifically, the central cusp appears shallower than a Bahcall--Wolf steady-state \citep{alexander&hopman09}, and recent measurements indicate $\gamma \approx 1.13$ \citep{schoedel+18}. The typical tidal separation radius of stellar binaries is larger than the single-star disruption radius by a factor $a_{\rm bin}/R_\star \sim \mathcal{O}(10)$, so we expect the critical radius for binary separation in the Milky Way to exist at several tens of parsecs (instead of the few parsecs expected for tidal disruption by Sgr A* in our Galactic Centre).

\begin{figure}
\includegraphics[width=120mm]{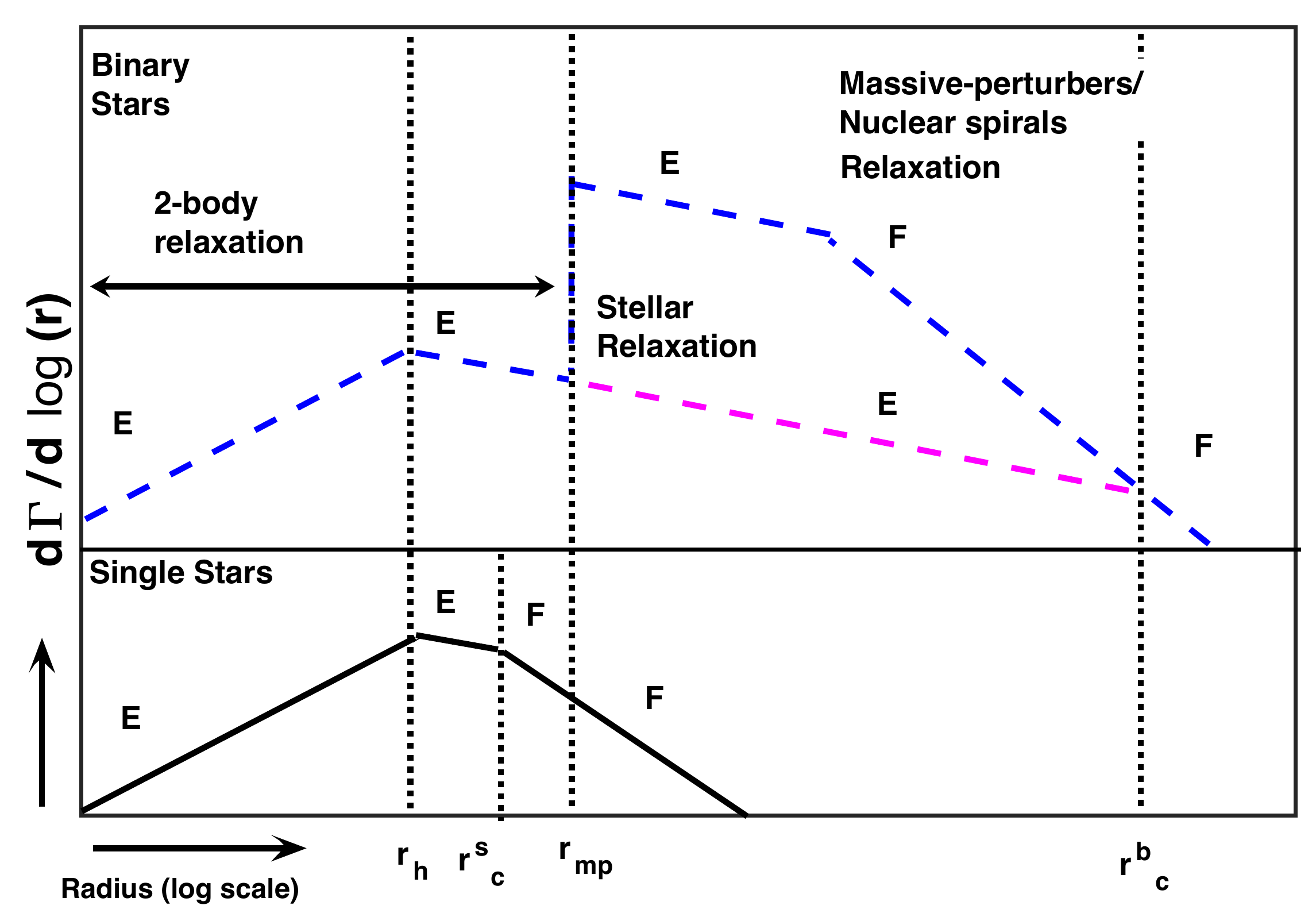}
\caption{A schematic representation of the local contribution to the loss-cone flux, for both binary and single stars. {\it Top}: the loss-cone flux for binary stars due to 2-body stellar relaxation (bottom dashed line) and due to massive perturbers (MP, top dashed line). Empty and full loss-cone regimes are denoted by ``E'' and ``F'', respectively. The radius $r_{\rm mp}$ corresponds to the region where such perturbers exist, likely just outside the innermost region of the nucleus. The critical radius for binaries and the sphere of influence radius are denoted respectively $r_{\rm c}^{\rm b}$ and $r_{\rm h}$. {\it Bottom}: the disruption rate of single stars. Here the presence of MPs coincides with the full loss cone regime and therefore little flux increase is expected. The critical radius for single stars is $r_{\rm c}^{\rm s} \sim r_{\rm h}$. Taken with permission from \citet{Perets+07a}, Fig. 3.}
\label{fig:empty-full-lc}
\end{figure}

As sketched in Fig. \ref{fig:empty-full-lc}, when two-body stellar relaxation is dominant and $\gamma < 9/4$, the {\it empty} loss cone (for binary separation) flux per unit bin of logarithmic radius increases outwards within the sphere of influence of the SMBH ($r_\mathrm{infl}$). On the other hand, it decreases outwards when $r>r_{\rm infl}$ \citep{Lightman&Shapiro77}. This implies that the loss cone flux for stellar binaries may peak around $r_\mathrm{\rm infl}$, with a long tail out to $\sim 100$~pc. As described earlier, the peak of the loss cone flux for single-star disruption comes from $\min(r_{\rm crit},r_{\rm infl})$, where typically $r_{\rm crit} \sim r_\mathrm{infl}$. Moreover, the empty loss cone flux is only logarithmically dependent on the loss cone size. These facts together might suggest that observational measurements of the rate of Galactic HVSs could be directly translated into constraints on TDE rates here\footnote{We note that since the counterparts of HVSs can be captured around the MBH, the distribution of such stars could also reflect the processes leading to, and the rates of, TDEs \citep{Perets&Gualandris2010}. }. This moment in time is especially propitious because of the ongoing data releases by the astrometric Galactic survey Gaia that have intensified searches for HVSs \citep[e.g.][]{marchetti+17,marchetti+19,boubert+18,bromley18}. Likewise, a comparison between rates of single and double TDE in other galaxies could provide an extra consistency check on the relationship between single and binary star loss cones. This comparison, however, requires more theoretical work in order to observationally distinguish these two types of transients \citep[e.g. by the presence of a precursor in a double TDE, as in][]{bonnerot&rossi19}. In general, both comparisons need to account for the fraction of binaries being separated versus disrupted \citep{bradnick+17}.

In principle, this is very exciting, but there are other physical ingredients -- irrelevant for single-star disruption -- that complicate the loss cone dynamics of binary stars.  The high stellar density inside the sphere of influence means that most soft binaries will not survive ``ionization'' (or ``evaporation'') from continuous gravitational interactions with field stars around them \citep{Perets+07a,hopman09,perets09,generozov18}. Hard binaries (progenitors of the fastest HVSs) can instead be driven to merger by magnetic braking.  The overall result is a binary to single ratio of less than $10\%$ at $1$ pc for low mass stars ($\leq 1 M_{\odot}$).  For more massive (but typically more rare) stars, the main sequence lifetime can be shorter than the evaporation timescale and the binary fraction would remain close to its birth value.  Evaporation might therefore drastically suppress the HVS rate from $r\lesssim r_{\rm infl}$, limiting our ability to directly calibrate TDE rates from HVS or double disruption rates, with the partial exception of massive stars. On the other hand, outside the sphere of influence, dynamical relaxation can be dominated by ``massive perturbers", such as giant molecular clouds, greatly shortening the relaxation timescale over that from two-body scatterings off stars. Since binary stars on loss cone orbits come from distances as far as $\sim 100 \times r_{\rm infl}$ and, unlike single stars, are in the empty loss cone regime (i.e. their flux {\it can} be increased), the presence of giant molecular clouds would enhance the HVS rate by a few orders of magnitude while leaving unaffected the predictions for TDEs \citep{Perets+07a,Hamers:2017}. In summary, constraining the TDE rate by observing HVSs in our Galaxy or double TDEs in external galaxies is in principle possible, but not straightforward.

\section{Applied Loss Cone Theory}
\label{sec:applied}

By changing the underlying density profile $\rho(r)$ or DF $f(\epsilon)$ in a spherical, isotropic galactic nucleus, a motivated theorist can tune the TDE rate to any desired value.  Asphericities and anisotropies offer further levers with which to change TDE rates.  In order to produce astrophysically realistic TDE rate estimates, the underlying galaxy model must, in some way, be calibrated off observations.  In this section, we will outline approximate but practical procedures for doing so. More specifically, we review how the theoretical loss cone physics of \S \ref{sec:losscone} may be combined with observations to make {\it empirically-calibrated} TDE rate estimates in nearby galaxies (\S \ref{sec:empirical}).  We will then examine the implications that past rate estimates along these lines have for distributions of TDE observables (\S \ref{sec:synthetic}).

\subsection{Simple Phase Space Modeling}
\label{sec:empirical}

In this subsection, we present a simple procedure for estimating TDE rates in individual galactic nuclei.  The key assumptions of this procedure, which was first developed by \citet{Magorrian&Tremaine99} and \citet{WangMerritt04}, are (i) spherical symmetry and (ii) nearly isotropic velocities, and this will be our starting point. 
Later on, we will also discuss more general prescriptions to account for geometrical asphericities or velocity anisotropies\footnote{The assumptions of spherical symmetry and quasi-isotropy are relaxed in \citet{Magorrian&Tremaine99}, but for brevity we focus primarily on the simplest case.}. 
  We begin with the observed 2D surface brightness profile, $I(R)$, which can be deprojected into a 3D luminosity density profile $j(r)$ using an Abel integral:
\begin{equation}
j(r) = \frac{-1}{\pi} \int_r^\infty \frac{{\rm d}I}{{\rm d} R} \frac{{\rm d} R}{\sqrt{R^2-r^2}}.
\end{equation}
Note that $R$ is a projected 2D radius and $r$ a 3D radius.  We then use a mass-to-light ratio, $\Upsilon$, to compute the 3D mass density $\rho(r) = \Upsilon j(r)$. 

This mass density profile can be used to compute other quantities of relevance to us, such as the period of a radial orbit, $T_{\rm orb}(E)$, the stellar mass enclosed $M_\star(r)$, and the gravitational potential $\Phi(r)$:
\begin{align}
    M_\star(r)=& \int_0^{r} 4\pi (r')^2 \rho(r'){\rm d}r' \\
    \Phi(r)=& -\frac{GM_\bullet}{r} - \frac{GM_\star(r)}{r} - 4\pi G \int_r^\infty \rho(r')r'{\rm d}r' \\
    T_{\rm orb}(E)=& 2\int_0^{r_\mathrm{max}} \frac{{\rm d}r}{\sqrt{2\big(E - \Phi(r)\big)}}.
\end{align}
Here we have also used the apocentre of a radial orbit, $r_\mathrm{max}(E)$.  Next, under the assumption of (nearly) isotropic velocities\footnote{If the stellar density profile $n(r)$ is too shallow, the DF $f(E)$ obtained from Eq. \ref{eq:Eddington} will have negative values, which is an unphysical outcome.  In the limit of a Kepler potential, the shallowest self-consistently isotropic power-law density profile is $n(r) \propto r^{-1/2}$; shallower density profiles require some degree of tangential anisotropy to remain positive-definite in $f(E,\mathcal R)$.}, we use Eddington's formula to compute a one-integral DF:
\begin{equation}
    f(E) = \frac{-1}{\pi^2 \sqrt{8}} \frac{{\rm d}}{{\rm d}E} \int^0_E \frac{{\rm d} n}{{\rm d}\Phi} \frac{{\rm d}\Phi}{\sqrt{\Phi - E}}. \label{eq:Eddington}
\end{equation}
In this integral, we convert stellar mass density to stellar number density $n(r)\equiv \rho(r) / \langle m_\star \rangle$.  The difference between these two mass functions is merely the average stellar mass $\langle m_\star \rangle$, which is determined by the stellar PDMF as in Eq. \ref{eq:PDMF}.  

With the isotropic DF $f(E)$ in hand, we may now apply the formalism of \S \ref{sec:relaxation} to obtain the TDE rate, $\dot{N}$, of the galaxy in question.  More specifically, we calculate the orbit-averaged diffusion coefficient $\mathcal{D}(E)$ using Eq. \ref{eq:diffusion_coefficient}, use this to calculate the diffusivity parameter $q(E)$ using Eq. \ref{eq:boundary_condition}, and then compute the loss cone flux $\mathcal{F}(E)$ as in Eq. \ref{eq:flux_general}.  Integrating $\mathcal{F}(E)$ across all energies $E$ gives the total TDE rate $\dot{N}$ (Eq. \ref{eq:total_rate}).  

In order to apply the above formalism to photometric observations of real galactic nuclei, we must consider a number of astrophysical uncertainties, including:
\begin{itemize}
    \item The functional form of $I(R)$: ideally, one would operate with nonparametric data, although some smoothing may be necessary to ensure positivity of $f(E)$.  However, in low-mass galaxies ($M_\bullet \sim 10^6 M_\odot$), the influence and critical radii are at best marginally resolved, so some degree of inward extrapolation in $I(R)$ is often necessary.  Past works have typically employed power-law fits to the innermost isophotes \citep{WangMerritt04, StoneMetzger16}, and the uncertainty produced by this extrapolation is in need of greater quantification.
    \item Choice of mass-to-light ratio $\Upsilon$: past efforts to predict TDE rates through dynamical modeling of large samples of galaxies have made crude estimates for $\Upsilon$.  For example, \citet{Magorrian&Tremaine99} employed the scaling relationship for V-band luminosities $\Upsilon_{\rm V} = 4.9 (M_\odot / L_\odot)(L_{\rm V} / 10^{10}L_\odot)^{0.18}$ \citep{Magorrian+98}.  Later works made virial estimates, starting with a galaxy's velocity dispersion $\sigma$, effective radius $R_{\rm eff}$, and luminosity $L$, then computing $\Upsilon_{\rm vir} = 2\sigma^2R_{\rm eff} / (GL)$ \citep{WangMerritt04, StoneMetzger16}.  Both of these methods make the large assumption that $\Upsilon$ is constant throughout the galaxy; a more self-consistent method would apply simple stellar population models to multiband photometry of the galactic nucleus to estimate $\Upsilon(R)$, as was done by \citet{StonevanVelzen16}.
    \item Choice of PDMF, ${\rm d}N/{\rm d}m_\star$: as we have seen, both the first and second moments of the PDMF enter into TDE rate calculations. Since the diffusion coefficients $\langle (\Delta \mathcal{R})^2 \rangle \propto \langle m_\star^2 \rangle$, the heaviest surviving stellar species will generally dominate rates of two-body relaxation.  For very young stellar populations this may be O stars, but for more typical stellar populations, it will be stellar-mass black holes.  Depending on the mass function of stellar-mass black holes considered \citep{Belczynski+10}, their inclusion in ${\rm d}N/{\rm d}m_\star$ will enhance TDE rates by factors of $\approx 1.3-4.9$ \citep{StoneMetzger16}.
    \item Determination of SMBH mass $M_\bullet$: all the studies described so far relied on galaxy scaling relations to estimate $M_\bullet$, due to scarcity of direct SMBH mass determinations. As such, they may be biased to various extents, depending on the assumed form of these relations, for which numerous and often incompatible versions have been produced over the last two decades.  See, for example, the comparison between different calibrations of the $M_\bullet-\sigma$ relation in the rate estimates of \citet{WangMerritt04}.
\end{itemize}
Aside from these astrophysical uncertainties, the simplifying assumptions introduce more fundamental limitations to the validity of this formalism.  So far, we have outlined a procedure for computing a one-integral, isotropic DF given only photometric information (i.e. the galaxy's surface brightness profile).  If additional kinematic information is available, a two-integral DF $f(E, L_{\rm z})$ may be computed instead \citep{Magorrian&Tremaine99}, which will self-consistently account for the impact of flattening and orbital anisotropy on TDE rates. 

\begin{figure}
\includegraphics[width=120mm]{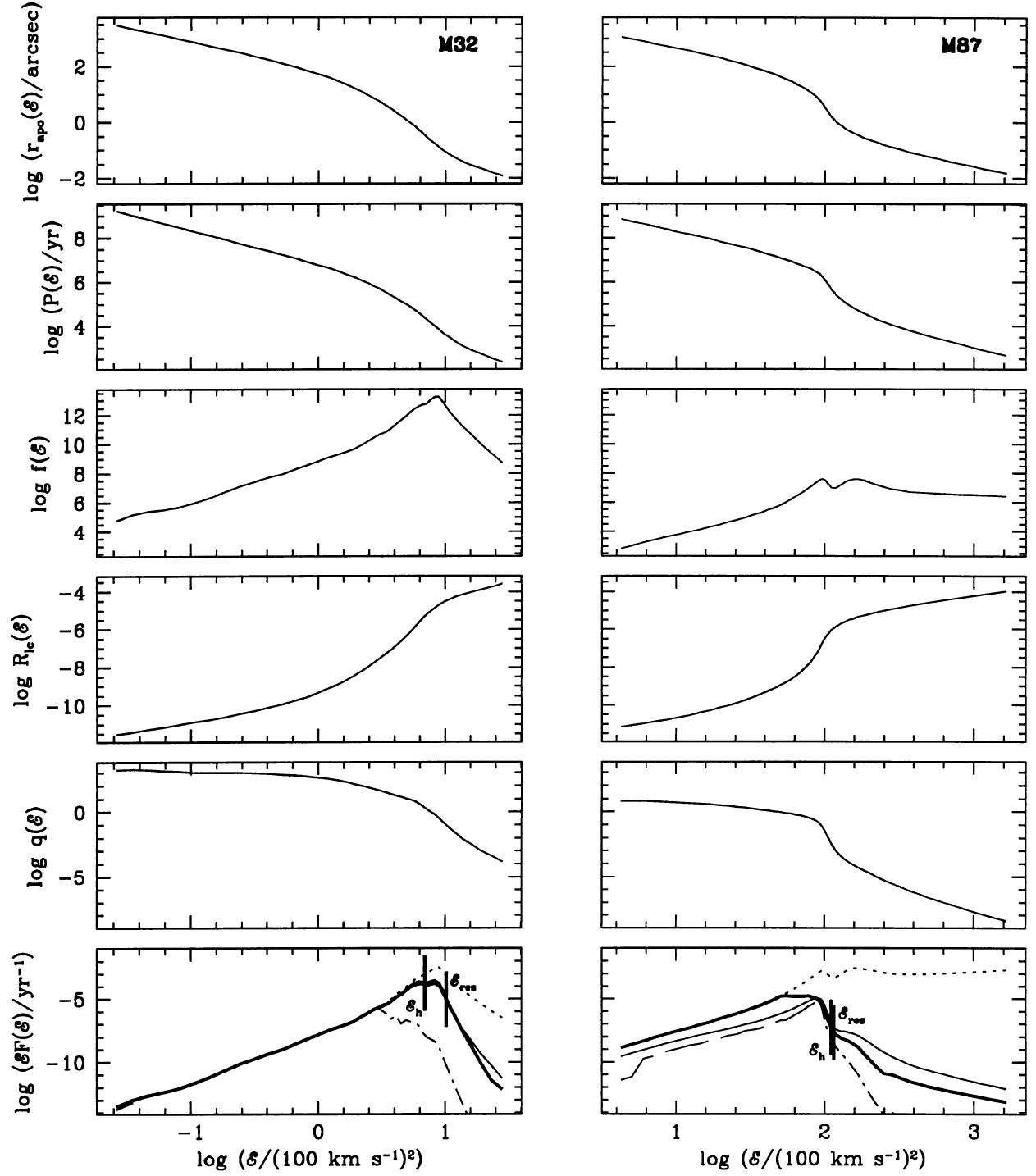}
\caption{Dynamically modeled photometry of two nearby galaxies, M32 (left) and M87 (right).  Figures show intermediate quantities and the final product of the modeling prescription described in \S \ref{sec:applied}, plotted in bins of (negative) specific orbital energy $\mathcal E \equiv -E > 0$.  The top two rows show apocentric radius $r_{\rm apo}(\mathcal E)$ and orbital period $P(\mathcal E)$ for radial orbits.  The next two rows show the one-integral DF $f(\mathcal E)$ (computed with an Eddington integral under the assumption of isotropic velocities) and the dimensionless size of the loss cone in angular momentum space, $\mathcal R_{\rm t}$ (labeled here as $\mathcal{R}_\mathrm{lc}$).  The second-to-last row shows the diffusivity parameter $q(\mathcal E)$ (for two-body loss cone repopulation), and the final row shows several different estimates for differential loss cone flux $\mathcal{F}(\mathcal E)$.  In the final row, the full loss cone flux is shown with a dotted line, the loss cone flux due to (spherical) two-body relaxation is a light solid line, the rate of draining of a full (axisymmetric) loss wedge is shown as a dot-dashed line, and the flux from a (axisymmetric) loss wedge being repopulated by two-body relaxation is shown as a dashed line.  The thick solid line represents the total TDE rate.  Taken with permission from \citet{Magorrian&Tremaine99}.}
\label{fig:modeledLW}
\end{figure}

The simple procedure outlined above has repeatedly been used to compute TDE rates in large galaxy samples, and at this point can be performed with the publicly available Fokker--Planck code \textsc{PhaseFlow} \citep{Vasiliev17}, as was done in, e.g. \citet{Pfister19}.  This basic procedure was first used, however, by \citet{SyerUlmer98}, who employed an even simpler formalism (one operating in coordinate space, not integral space) to a sample of 25 galaxies with surface brightness profiles $I(R)$ taken from \citet{Byun+96} and SMBH masses taken from \citet{Magorrian+98}.  The computed TDE rates were very low, with $10^{-7}~{\rm yr}^{-1} \lesssim \dot{N} \lesssim 10^{-4}~{\rm yr}^{-1}$ across most of the sample, although only 6 out of the 25 galaxies had SMBHs smaller than the Newtonian Hills mass.  Almost simultaneously, \citet{Magorrian&Tremaine99} used a more sophisticated, two-integral version of the loss cone formalism to analyze a sample of 29 galaxies, taking into account multiple sources of loss cone flux: standard two-body relaxation, the draining of a loss wedge region in an axisymmetric potential, and two-body repopulation of the loss wedge.  Two-integral DFs $f(E, L_{\rm z})$ were taken from \citet{Magorrian+98}.  This work found relatively low rates of disruption from two-body relaxation (albeit a factor of a few higher than what was found in \citealt{SyerUlmer98}), but once again, focused primarily on the largest galactic nuclei, with only 3 out of the 29 galaxies possessing SMBHs below the Schwarzschild Hills mass.  This study was the first to point out that the observed dichotomy in nuclear density profiles $n(r)$ -- between steeply declining ``cusp'' galaxies and relatively shallow ``core'' galaxies \citep{Lauer+95} -- implies that the highest TDE rates will occur in lower-mass, ``cuspy'' galaxies.  We reproduce many of the derived quantities, such as $q(E)$ and $\mathcal{F}(E)$, from \citet{Magorrian&Tremaine99} in Fig. \ref{fig:modeledLW}.

\begin{figure}
\includegraphics[width=120mm]{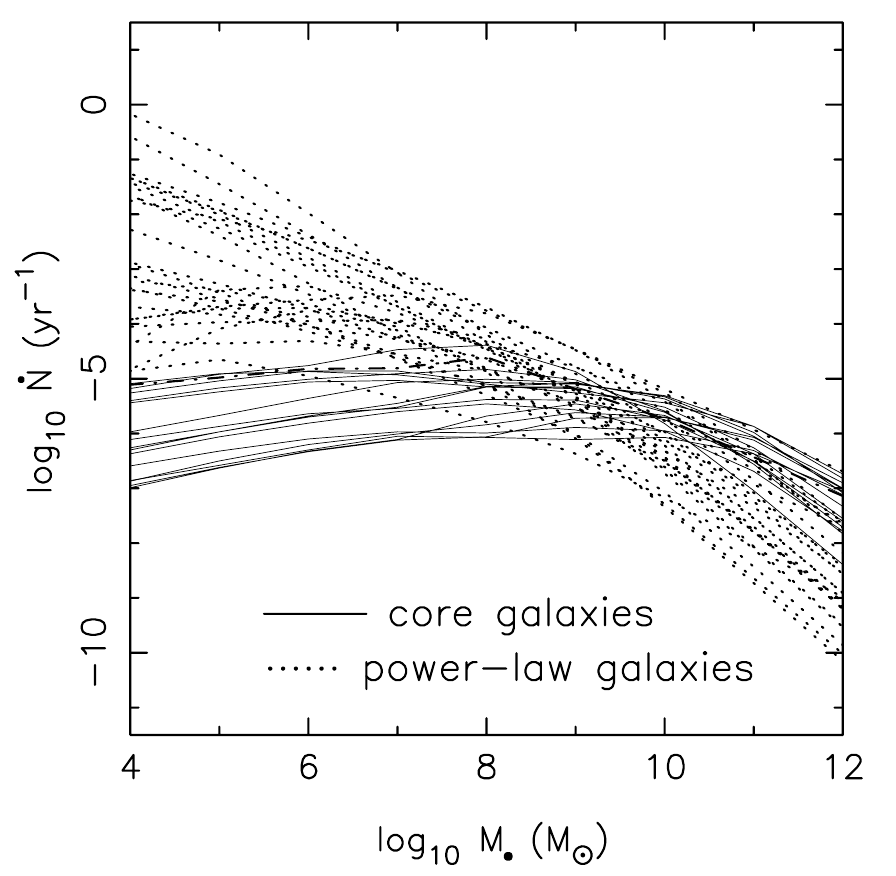}
\caption{The dependence of stellar consumption rates $\dot{N}$ on assumed SMBH mass $M_\bullet$ in a sample of 41 galaxies modeled with {\it HST} photometry (note that main-sequence stars will not produce TDEs for $M_\bullet > M_{\rm H}\sim 10^8\,M_\odot$). The rates were computed assuming spherical symmetry and two-body relaxation, using the prescriptions of \S\ref{sec:applied}, but allowing SMBH mass to float.  An interesting dichotomy emerges: for cusp (``power-law'') stellar distributions, TDE rates increase with decreasing $M_\bullet$, while TDE rates are roughly independent of $M_\bullet$ in flatter, core-like nuclei.  Taken with permission from \citet{WangMerritt04}.}
\label{fig:coreCusp}
\end{figure}

A larger sample of 41 galaxies was modeled using one-integral DFs $f(E)$ by \citet{WangMerritt04}, following almost exactly the procedure of this subsection.  These authors estimated SMBH masses using the $M_\bullet -\sigma$ relationship of \citet{MerrittFerrarese01}.  This scaling relationship predicts systematically lower $M_\bullet$ values than that of \citet{Magorrian+98}, and as a result this study found much higher TDE rates, with typical $\dot{N} \sim 10^{-4}~{\rm yr}^{-1}$ in sub-Hills mass nuclei.  This sample, which took $I(R)$ profiles from \citet{Faber+97}, was more applicable to realistic TDE hosts than past studies, with 21 out of 41 galaxies possessing SMBHs below the Schwarzschild Hills mass.  \citet{WangMerritt04} established much more firmly that the volumetric rate of TDEs, $\dot{n}$, should be dominated by the lowest-mass galaxy bin with a high SMBH occupation fraction.  The reasons for this are (i) the greater abundance of low-mass galaxies in the Universe, (ii) the empirical preference of low-mass galaxies to have steep central stellar density cusps, which produce higher per-galaxy TDE rates $\dot{N}$, and (iii) the anticorrelation between $\dot{N}$ and SMBH mass $M_\bullet$ in a cuspy profile (see Fig. \ref{fig:coreCusp} for the identification of this trend in \citealt{WangMerritt04}).

\begin{figure}
\includegraphics[width=120mm]{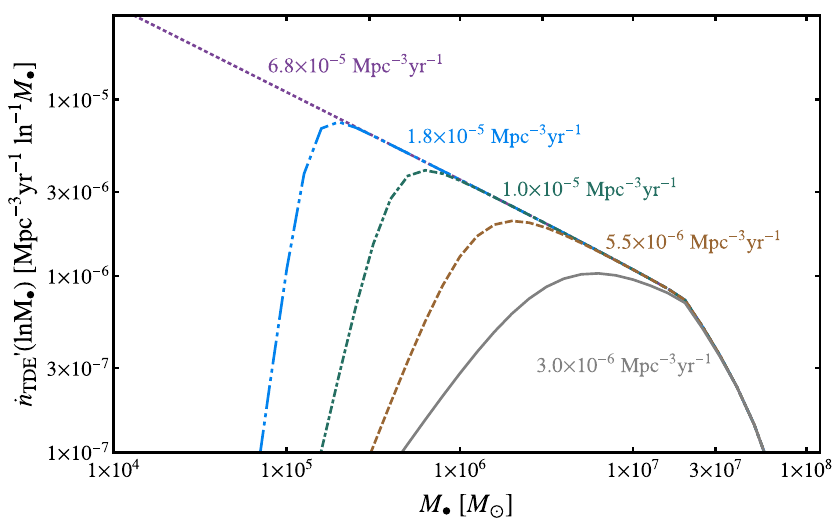}
\caption{The volumetric TDE rate per dex, ${\rm d}\dot{n} / {\rm d}\ln M_\bullet$, plotted as a function of SMBH mass $M_\bullet$.  Per-galaxy TDE rates are estimated by power-law fits to the results of a large, dynamically modeled galaxy sample, and are then coupled to a Schechter function that is populated with SMBHs using the $M_\bullet-\sigma$ relation and five different models for the SMBH occupation fraction in low-mass nuclei.  Specifically, the SMBH occupation fraction is assumed to resemble a step function, with SMBHs largely absent below a cutoff host stellar mass $M_{\rm cut}/M_\odot$ of $10^9$ (grey solid), $10^{8.5}$ (brown dashed), $10^8$ (green dot-dashed), $10^{7.5}$ (blue dot-dot-dashed), and a final scenario in which every dwarf galaxy has massive black holes down to a limiting mass of $M_\bullet=10^4 M_\odot$ (purple dotted).  Volumetric TDE rates are dominated by the smallest galaxies that typically host SMBHs.  Mass-integrated rates $\dot{n}$ are labeled for each of these five models.   Taken with permission from \citet{StoneMetzger16}.}
\label{fig:modeledRate}
\end{figure}

More recently, empirically calibrated TDE rates were re-examined by \citet{StoneMetzger16}, who used the one-integral formalism of this section to model a sample of 144 galaxies \citep{Lauer+07a, Lauer+07b}, of which 42 contain SMBHs with $M_\bullet$ below the Schwarzschild Hills mass.  This work was the first to estimate the fraction, $f_{\rm pin}$, of TDEs from a given galaxy in the $q(E)>1$ (pinhole) regime of disruption.  Empirically, $f_{\rm pin}$ is $\mathcal{O}(1)$ in galaxies with $M_\bullet \lesssim 10^7 M_\odot$, but exhibits broad scatter ($10^{-2} < f_{\rm pin} < 1$) at higher SMBH masses.  In any given bin of $M_\bullet$, the pinhole fraction is much higher in core galaxies than in cusp galaxies.  \citet{StoneMetzger16} reproduced earlier findings that per-galaxy TDE rates $\dot{N}$ are highest in smaller galaxies, and fit power-laws to their sample of dynamically modeled nuclei:
\begin{equation}
    \dot{N} = \dot{N}_8 \left(\frac{M_\bullet}{10^8 M_\odot} \right)^B. \label{eq:empiricalRate}
\end{equation}
Here $\dot{N}_8 = 2.9 \times 10^{-5}~{\rm yr}$ and $B=-0.404$ when considering the entire sample of 144 modeled galaxies.  For a subsample of exclusively core (cusp) galactic nuclei, $\dot{N}_8 = 1.2 \times 10^{-5}~{\rm yr}$ and $B=-0.247$ ($\dot{N}_8 = 6.5 \times 10^{-5}~{\rm yr}$ and $B=-0.223$).  These results were combined with a Schechter function, the Faber--Jackson law, a recent calibration of the $M_\bullet - \sigma$ relation \citep{McConnell&Ma13}, and various parametrizations of the SMBH occuption fraction to estimate the volumetric TDE rate, $\dot{n}(M_\bullet)$, which is presented here in Fig. \ref{fig:modeledRate}.  These results indicate that a volume-complete sample of TDEs will be a powerful probe of the unknown bottom end of the SMBH mass function.

Several astrophysical uncertainties apply to existing empirically-calibrated TDE rate estimates, as we have listed above.  Aside from these caveats, it is notable that no loss cone calculations have yet been performed using the richer dynamical information available from Jeans or Schwarzschild modeling of a large sample of galaxies with (i) resolved kinematics and (ii) $M_\bullet < M_{\rm H}$.  The pioneering work of \citet{Magorrian&Tremaine99} established the basic theoretical framework for this effort, but was applied almost exclusively to SMBHs too large to produce luminous TDEs.  Applying loss cone theory to galaxies with directly measured SMBH masses and anisotropy profiles would enhance the precision of TDE rate estimates substantially, and is a logical next step forward.

\subsection{Synthetic Observables}
\label{sec:synthetic}

In the previous subsection, we surveyed existing efforts to empirically calibrate volumetric ($\dot{n}$) and per-galaxy ($\dot{N}$) TDE rates.  However, the observable properties of TDE flares are not uniform, and can depend strongly on event parameters such as SMBH mass $M_\bullet$ or penetration parameter $\beta$.  If one knows the differential distribution of TDE rates with respect to controlling parameters such as these, then it is straightforward to compute distributions of observable parameters of interest.  Some of these parameters, such as the peak mass fallback rate $\dot{M}_{\rm peak}$ (or its Eddington ratio, $\dot{M}_{\rm peak}/\dot{M}_{\rm Edd}$), are very well-understood theoretically.  Others, such as peak optical luminosity, are much less well-understood.  Much more detailed theoretical treatments of TDE observables are offered in later Chapters of this Volume.  In this section, we do not provide detailed models for TDE observables, but advise the interested reader to consult the \disrupchap{}, \flowchap{}, \diskchap{}, and \emischap{}.

Dynamical modeling of nearby galactic nuclei has already allowed us to estimate the $M_\bullet$-dependence of TDE rates, ${\rm d}\dot{n}/{\rm d}M_\bullet$ (e.g. Fig. \ref{fig:modeledRate}).  The $\beta$-dependence, ${\rm d}\dot{n}/{\rm d}\beta$, is a little more complicated, but at least in the limit of spherical loss cone repopulation, it can be estimated approximately.  In the pinhole regime ($q \gg 1$), the differential distribution of penetration parameters goes as ${\rm d}\dot{n}/{\rm d}\beta \propto \beta^{-2}$, while in the diffusive regime ($q \ll 1$), $\beta \approx 1$, and higher $\beta$ values are exponentially suppressed (see also the discussion surrounding Eq. \ref{eq:Ndot_partial}).  \citet{StoneMetzger16} find that low-mass galactic nuclei ($M_\bullet \lesssim 10^7 M_\odot$) typically produce most of their TDEs from the pinhole regime, while there is much more scatter in the ``pinhole fraction'' of high-mass galactic nuclei (although it is rarely less than $10\%$, for $M_\bullet < M_{\rm H}$).  For any given bin of $M_\bullet$, the pinhole fraction is lower in cusp and higher in core nuclei.

\begin{figure}
\includegraphics[width=120mm]{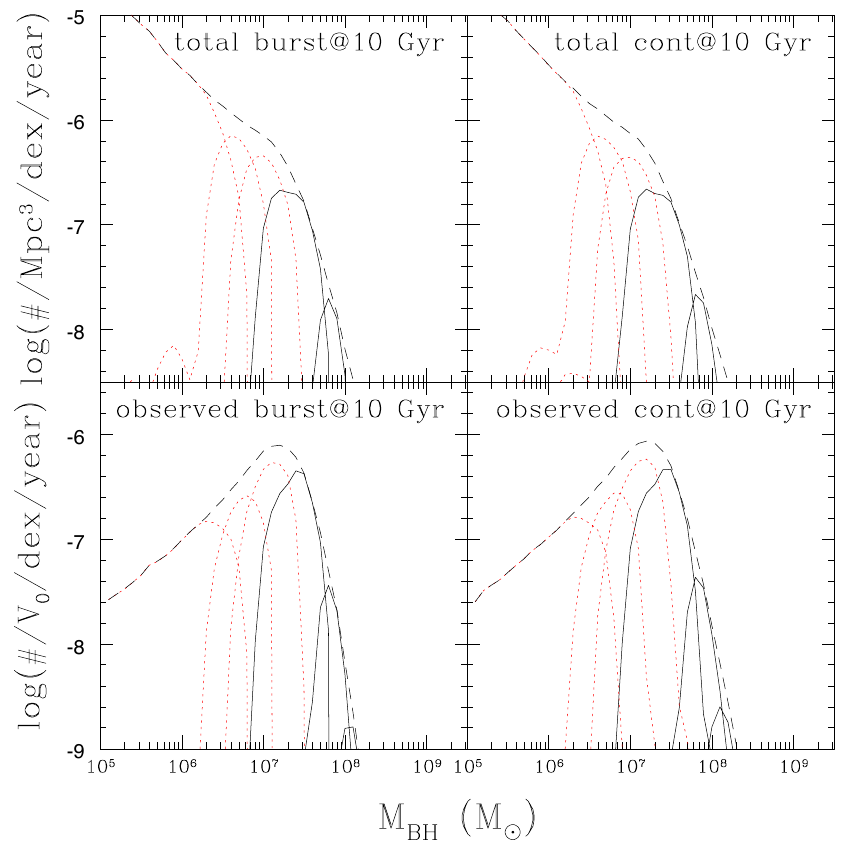}
\caption{The differential distribution of volumetric TDE rates (top panels) ${\rm d}\dot{n}/{\rm d}\ln M_\bullet$ and observed rates per dex ${\rm d}\dot{\mathcal{N}}/{\rm d} \ln M_\bullet$ (bottom panels).  Because these TDE rates are integrated over PDMFs, they depend on assumed star formation histories; the left panels assume a single burst of star formation 10 Gyr ago, while the right panels assume a constant star formation rate over a Hubble time.  Dashed black lines show total TDE rates ${\rm d}\dot{n}/{\rm d}\ln M_\bullet$ and ${\rm d}\dot{\mathcal{N}}/{\rm d} \ln M_\bullet$, while the other lines break up the rates into bins of $\dot{m}_{\rm peak} \equiv \dot{M}_{\rm peak}/\dot{M}_{\rm Edd}$.  The red dotted lines correspond to super-Eddington peak fallback rate regimes of $10^{0.0}<\dot{m}_{\rm peak} <10^{0.5}$, $10^{0.5}< \dot{m}_{\rm peak} <10^{1.0}$, and $10^{1.0}> \dot{m}_{\rm peak}$, while the black solid lines correspond to sub-Eddington peak fallback rate regimes of $10^{-0.5}< \dot{m}_{\rm peak} <10^{0.0}$, $10^{-1.0}< \dot{m}_{\rm peak} <10^{-0.5}$, $10^{-1.5}< \dot{m}_{\rm peak} <10^{-1.0}$, and $ \dot{m}_{\rm peak} <10^{-1.5}$.  In translating from intrinsic rates to observed rates, the calculation assumes a flux-limited survey and that TDE peak luminosities are determined by either the peak mass fallback rate or the Eddington-limited accretion rate, whichever is smaller.  Taken with permission from \citet{Kochanek16}.}
\label{fig:modeledDemographics}
\end{figure}

The distribution of fallback times $t_{\rm fall}$ and the closely related peak fallback rate $\dot{M}_{\rm peak}$ may be estimated using the approximate analytic formulae \citep{Rees88, Stone+13}:
\begin{align}
    t_{\rm fall} \approx& 3.5\times 10^6~{\rm s}~\left( \frac{M_\bullet}{10^6 M_\odot} \right)^{1/2} \left( \frac{M_\star}{ M_\odot} \right)^{-1} \left( \frac{R_\star}{R_\odot} \right)^{3/2} \\
    \frac{\dot{M}_{\rm peak}}{\dot{M}_{\rm Edd}} \approx & 130  \left( \frac{\eta}{0.1} \right) \left( \frac{M_\bullet}{10^6 M_\odot} \right)^{-3/2} \left( \frac{M_\star}{ M_\odot} \right)^{2} \left( \frac{R_\star}{R_\odot} \right)^{-3/2}.
\end{align}
Here we have defined the Eddington-limited mass accretion rate, $\dot{M}_{\rm Edd} \equiv L_{\rm Edd}/(\eta c^2)$, in terms of a radiative efficiency $\eta<1$ and the Eddington luminosity $L_{\rm Edd} \approx 1.5 \times 10^{46}~{\rm erg~s}^{-1}~(M_\bullet / 10^8 M_\odot)$.  While these observables can be computed more precisely using hydrodynamical disruption simulations \citep{Guillochon&RamirezRuiz13}, these analytic expressions were combined with TDE rate computations to predict distributions of peak fallback rates and rise times in \citet{StoneMetzger16} and \citet{Kochanek16}.  We present the results of the latter paper in Fig. \ref{fig:modeledDemographics}, which plots ${\rm d}\dot{n}/{\rm d}\ln M_\bullet$ in different bins of peak Eddington ratio $\dot{M}_{\rm peak}/\dot{M}_{\rm Edd}$.  These distributions (which are integrated over different stellar PDMFs, corresponding to different star formation histories) highlight that for $M_\bullet \lesssim 10^7 M_\odot$, the large majority of TDEs have initially super-Eddington fallback rates, while the reverse is true for $M_\bullet \gtrsim 10^{7.5}M_\odot$.  

Fig. \ref{fig:modeledDemographics} also presents the TDE rates that will be {\it observed} in a flux-limited survey, $\dot{\mathcal{N}}(M_\bullet)$.  These results assume a simple model relating peak luminosity to $M_\bullet$; more complicated models are explored in \citet{StoneMetzger16} and \citet{Mageshwaran&Mangalam15}.  Generally, if super-Eddington fallback rates can be translated linearly into super-Eddington luminosities, then $\dot{\mathcal{N}}(M_\bullet)$ is dominated by the smallest values of $M_\bullet$ that exist with a high occupation fraction.  Conversely, if TDE emission mechanisms are Eddington-limited, then $\dot{\mathcal{N}}(M_\bullet)$ is dominated by $M_\bullet \sim 10^7 M_\odot$, the characteristic SMBH mass where $\dot{M}_{\rm peak}/\dot{M}_{\rm Edd} \sim 1$.  A more recent estimate of the volumetric detection rate, using a simple phenomenological model for TDE luminosities, is available in \citet{Thorp+18}, which also compares standard TDE rates to those from binary SMBHs.

Other notable conclusions of efforts to produce synthetic observable distributions include: (i) regardless of star formation history, both intrinsic and observed TDE rates are dominated by the bottom end of the stellar IMF, except when $M_\bullet \sim M_{\rm H}$ \citep{StoneMetzger16, Kochanek16}; (ii) under the assumption that galactic nuclei have similar dynamical properties at all redshifts\footnote{This assumption is unlikely to be generically true.  The clearest caveat here concerns the strong preference among observed TDE flares to reside in rare E+A and, more generally, post-starburst galaxies (see discussion in \S \ref{sec:poststarburst}).  Because E+A and post-starburst galaxies make up very small fractions of the low-$z$ galaxy population ($\sim 0.2\%$ and $2.3\%$, respectively \citep{French+16}, this implies the presence of unusual stellar dynamics enhancing TDE rates in these galaxies by at least an order of magnitude.  However, the fraction of all galaxies that have a post-starburst nature increases steeply as a function of redshift.  For example, going from $z\approx 0.5$ to $z\approx 2$ increases the fraction of post-starburst galaxies by a factor of $\approx 5$ \citep{Wild+16}, suggesting that at high $z$, the decline in $\dot{n}$ due to the decreasing volume density of SMBHs may be overwhelmed by the growing abundance of this rare galaxy type.}, comoving volumetric TDE rates will decline by a factor $\approx 10$ from their $z=0$ value as one moves to $z=2$ \citep{Kochanek16}, largely due to the decreasing volume density of SMBHs at high $z$ \citep{Shankar+09}; (iii) k-corrections generally aid the detectability of TDEs at cosmological distances \citep{StrubbeQuataert09}.  Unfortunately, the current lack of accurate, first-principles models for TDE emission mechanisms has, at the time of writing, hampered the calculation of more specific distributions of observables (e.g. emission line strengths, X-ray to optical ratios, etc.).

So far we have only discussed the observable properties of main sequence TDEs, but detailed distributions of red giant disruption properties (such as stage of post-main sequence evolution) are available in \citet{MacLeod+12}.  Synthetic distributions of fallback times for helium white dwarf and brown dwarf disruptions were, likewise, computed in \citet{LawSmith+17b}.

\section{Comparing to Observed Rates of Tidal Disruption}
\label{sec:comparison}
For over two decades, tidal disruptions were studied from purely theoretical grounds.  This situation changed in the 1990s with the discovery, in soft X-ray energies, of the first strong TDE candidates \citep[][see also the \xraychap{}]{Bade+96, Komossa&Greiner99}.  Later, wide-field ultraviolet \citep{Gezari+06} and optical \citep[][see also the \optchap{}]{vanVelzen+11} surveys discovered other classes of TDE candidates.  At the time of writing, the current rate of TDE discovery stands at $\approx 1-3$ new TDE candidates per year, and a few dozen TDE candidates are known (see e.g. the compilations of \citealt{Auchettl+17, Hung+17}).  New and upcoming time domain surveys are poised to expand our sample by orders of magnitude \citep{vanVelzen+11, Khabibullin+14, Mageshwaran&Mangalam15}.  Clearly, it is important to assess what scientific goals can be achieved with the statistical analysis of large near-future TDE samples.  A key component of any such analysis is understanding the rate of tidal disruption.  

In this section, we compare the theoretical rate estimates of \S \ref{sec:applied} to those inferred from the existing sample of tidal disruption flares.  This type of comparison has, so far, found two interesting puzzles, which we outline in \S \ref{sec:observationalRates}.  First, there may be a broad rate discrepancy, with fewer TDEs detected than are predicted by stellar dynamics \citep{StoneMetzger16}.  Second, there is a more specific and definitive discrepancy related to the {\it overproduction} of observed TDEs in a rare galaxy subclass: so-called ``E+A,'' or post-starburst, galaxies \citep{Arcavi+14}.  We discuss these puzzles in \S \ref{sec:discrepancies}.

\subsection{Observationally Inferred Rates}
\label{sec:observationalRates}
The first TDE candidates were discovered via soft X-ray emission \citep{Bade+96}, and the first observational rate inferences were based off of these events.  The pioneering work of \citet{Donley+02} analyzed three TDE candidates from the {\it ROSAT} All-Sky Survey and inferred a per-galaxy disruption rate $\dot{N} \approx 9 \times 10^{-6}~{\rm yr}^{-1}$.  Subsequent analysis of two TDE candidates from the {\it XMM-Newton} slew survey inferred a volumetric TDE rate $\dot{n} \approx 5.4\times 10^{-6}~{\rm yr}^{-1}~{\rm Mpc}^{-3}$, and a per-galaxy rate $\dot{N} \approx 2.3 \times 10^{-4}~{\rm yr}^{-1}$ \citep{Esquej+08}.  Later, one soft X-ray TDE candidate was found by reanalysis of archival {\it Chandra} and {\it XMM-Newton} observations of galaxy clusters \citep{Maksym+10}; three more were discovered by cross-correlating {\it ROSAT} archives with serendipitous {\it XMM-Newton} pointings \citep{KhabibullinSazonov14}.  The per-galaxy event rates computed from these samples were $\dot{N} \approx 1.2 \times 10^{-4}~{\rm yr}^{-1}$ and $\dot{N} \approx 3 \times 10^{-5}~{\rm yr}^{-1}$, respectively.  The sample of \citet{KhabibullinSazonov14} implies a volumetric event rate $\dot{n} \approx 4-8\times 10^{-7}~{\rm yr}^{-1}~{\rm Mpc}^{-3}$.

The range of inferred X-ray TDE rates spans roughly 1.5 orders of magnitude.  This spread can be attributed to a number of factors.  Obviously, small-number statistics play a role.  Due to the poor temporal resolution of all wide-field soft X-ray surveys to date, these samples are both flux- and cadence-limited, and rate inference therefore requires assumptions about (i) the X-ray luminosity function of TDEs, and (ii) the characteristic decay time of TDEs.  With these assumptions, one may estimate survey completeness as a function of source distance, and therefore infer a volumetric event rate $\dot{n}$.  As a simplified but illustrative example, consider a high-cadence survey operating at low redshifts for a duration $T$, covering an angular area $\Delta \Omega$.  In this limit, the number of detected TDEs
\begin{equation}
    N_{\rm TDE} = \int \frac{{\rm d}\dot{n}}{{\rm d}L} \times \frac{4\pi}{3}D_{\rm max}^3(L) \times \frac{\Delta \Omega}{4\pi} \times T {\rm d}L.
\end{equation}
Here $D_{\rm max}(L)$ is the maximum distance out to which a TDE can be detected in a flux-limited survey, and  ${\rm d}\dot{n}/{\rm d}L$ is the volumetric TDE luminosity function \citep{Hung+18}.  By assuming approximate functional forms for the luminosity function, one may use a measured $N_{\rm TDE}$ to estimate $\dot{n}\equiv \int({\rm d}\dot{n}/{\rm d}L){\rm d}L$.  Translating this $\dot{n}$ into a per-galaxy rate $\dot{N}$ requires further assumptions about {\it which} galaxies may produce TDEs (e.g. down to what limiting mass do dwarf galaxies still possess SMBHs?).  Given these many uncertainties, it is perhaps unsurprising to see substantial scatter in observationally inferred X-ray event rates.  

More recently, the discovery rate of TDEs has been dominated by wide-field optical/UV surveys.  The first TDE candidates discovered from their thermal UV emission were found with the {\it GALEX} satellite \citep{Gezari+06}.  The rate of TDE discovery by {\it GALEX} was found to be consistent \citep{Gezari+08} with the empirical predictions of \citet{WangMerritt04}.  Soon afterwards, TDE candidates began to be discovered through their thermal optical emission -- at first archivally \citep{vanVelzen+11}, and soon afterwards in real-time surveys \citep{Gezari+12}.  As the optically-selected TDE sample has grown, several groups have attempted to infer the TDE rates $\dot{N}$ and $\dot{n}$ associated with optically-bright flares.  One advantage that rate inferences from optically-selected TDEs have over those from X-ray-selected TDEs is the much higher cadence of optical time-domain surveys.  This means that rate estimates will be less sensitive to assumptions about the temporal evolution of TDE light curves.

Initially, rates estimated from optically-selected TDEs were low.  For example, the discovery of two TDE candidates in SDSS archival data was used to compute a per-galaxy rate $\dot{N} \approx (1.5-2.0)^{+2.7}_{-1.3}\times 10^{-5}~{\rm yr}^{-1}$, and a volumetric rate $\dot{n} \approx (4-8) \times 10^{-8\pm 0.4}~{\rm yr}^{-1}~{\rm Mpc}^{-3}$ \citep{vanVelzenFarrar14}\footnote{Note that in the results of \citet{vanVelzenFarrar14}, statistical uncertainties are denoted in superscript/subscript error ranges, while systematic uncertainties (associated with the uncertain choice of model light curve used to back out true rates from flux-limited samples) are denoted in prefactor error ranges.}.  
The later discovery of two TDE candidates by the ASAS-SN survey implied a somewhat higher per-galaxy rate of $\dot{N} \approx 4.1_{-1.9}^{+12.9} \times 10^{-5}~{\rm yr}^{-1}$ \citep{Holoien+16}.  More recent calculations that use an {\it empirical} TDE luminosity function find higher event rates.  For example, \citet{vanVelzen18} analyze a sample of thirteen TDE candidates found in optical/UV surveys and find $\dot{n} = (8 \pm 4) \times 10^{-7}~{\rm Mpc}^{-3}~{\rm yr}^{-1}$, consistent with a per-galaxy rate $\dot{N} \approx 1\times 10^{-4}~{\rm yr}^{-1}$.  Comparably high-per galaxy rates are also inferred from two TDE candidates found by iPTF ($\dot{N} \approx 1.7^{+2.9}_{-1.3}\times 10^{-4}~{\rm yr}^{-1}$; \citealt{Hung+18}).

One particularly intriguing observational finding is an apparent overabundance of TDE candidates in quiescent galaxies with no ongoing star formation and prominent Balmer absorption features.  These absorption features are generally produced by a large population of young A stars, indicating that this type of galaxy (in its extreme form, sometimes known as an ``E+A'' or ``K+A'' galaxy) recently underwent a major star formation episode, that has since ceased.  Although this post-starburst preference was first noted among optically-selected TDEs \citep{Arcavi+14}, it appears to be generic across most classes of candidate TDE flares \citep{Graur+18}.  While the {\it absolute} rate density of TDEs in post-starburst galaxies, $\dot{n}_{\rm PS}$, is subject to the same uncertainties discussed above, the relative rate enhancement factor $\mathscr{R} \equiv \dot{n}_{\rm PS} / \dot{n}$, may be computed with much less uncertainty (provided that the TDE luminosity function is nearly the same in normal and post-starburst galaxies).

Interestingly, the rate enhancement seems to depend strongly on the strength of the Balmer absorption features.  The first detailed analysis of post-starburst TDE hosts found that, in the most Balmer-strong post-starbursts (classical E+As), $\mathscr{R} = 190_{-100}^{+191}$, while in less extreme Balmer-strong post-starbursts, $\mathscr{R} = 33_{-11}^{+7}$ \citep[][for a sample of 8 TDE hosts]{French+16}.  
The strength of the Balmer absorption feature represents a weighted combination of post-starburst age, fraction of the galaxy starlight produced in the starburst, and the detailed shape of star formation history during and after the starburst \citep{French+17}.  A subsequent analysis \citep{Graur+18} of a larger TDE host sample found somewhat smaller but still significant rate enhancements ($\mathscr{R} = 35_{-17}^{+21}$ and $\mathscr{R} = 18_{-7}^{+8}$ for the same categories of Balmer-line strength).  The \citet{Graur+18} rate enhancements were computed from a larger but more heterogeneous sample of 33 TDE host galaxies, and vary significantly if one examines different sub-samples.  TDE host galaxies display some other peculiarities as well, such as an unusually high surface brightness on $\sim \rm kpc$ scales \citep{LawSmith+17}.  The properties of TDE hosts are not yet fully understood, and our empirical knowledge of this subject is evolving rapidly.  The interested reader is advised to consult the \hostchap{} for a more comprehensive picture.

As TDE samples grow, it will become possible to make more refined comparisons between theory and observation.  For example, in Fig. \ref{fig:modeledRate}, we can see that the predicted, volumetric TDE rate is expected to significantly depend on $M_\bullet$.  Observational evidence testing this prediction is mixed, although still limited by small-number statistics.  Early work by \citet{StoneMetzger16} used galaxy scaling relations to find that an optical/UV selected TDE sample (consisting of 11) had a mass distribution sharply peaked near $10^{6.8}M_\odot$, with little evidence for optical/UV TDE hosts possessing SMBHs below $10^6 M_\odot$.  Subsequent work by \citet{Wevers+19} improved on this by using a larger TDE sample (15 flares) and also using a more homogenous set of SMBH mass estimates.  The \citet{Wevers+19} histogram shows a more broadly peaked distribution centered on somewhat lower masses ($\approx 10^{6.2}M_\odot$), but with a qualitatively similar shape.  In contrast, both of these analyses find a much broader distribution of host masses for TDEs selected through soft X-ray emission. In comparison to the optical/UV selected sample, the distribution of inferred SMBH masses in X-ray TDEs is less sharply peaked, with a greater fraction of these events occurring in galaxies with very high-mass ($M_\bullet \gtrsim 10^{7.5}M_\odot$) or very low-mass ($M_\bullet \lesssim 10^{5}M_\odot$) SMBHs.  

Neither of these analyses, however, accounted for survey selection effects, such as volume-correcting flux-limited samples. To date, the most thorough effort to back out volumetric rates $\dot{n}(M_\bullet)$ from observations comes from \citet{vanVelzen18}, who find an effectively constant TDE rate for bins of SMBH mass between $10^6 M_\odot$ and $10^{7.5}M_\odot$.  At present, the two biggest obstacles to drawing inferences about the bottom end of the SMBH mass function from observed TDEs are (i) small, inhomogenous TDE samples and (ii) the uncertain luminosity function of TDEs, particularly those from smaller SMBHs where Eddington considerations may be more important.

\subsection{Rate Discrepancies}
\label{sec:discrepancies}
As we saw in \S \ref{sec:applied}, there are many uncertainties associated with dynamical modeling of nearby galactic nuclei.  However, a few general conclusions stand out:
\begin{enumerate}
    \item TDE rates $\dot{N}$ are generally higher in smaller galaxies, if we consider only two-body relaxation in spherical potentials\footnote{The situation becomes more complicated if galactic nuclei are significantly triaxial, in which case larger galaxies may have larger individual TDE rates $\dot{N}$.  From an observational point of view, the prevalence of nuclear triaxiality remains uncertain.}.  This is primarily due to the higher central densities and steeper density profiles associated with low-mass galactic nuclei. 
    \item The differential volumetric TDE rate ${\rm d} \dot{n}/{\rm d}\ln M_\bullet$ varies across host galaxy mass, but the integrated rate $\dot{n}$ is likely dominated by the lowest-mass range galaxies that have a high SMBH occupation fraction. 
    \item Averaged over the set of galaxies with a high black hole occupation fraction, per-galaxy TDE rates are $\dot{N} \gtrsim 1\times 10^{-4}~{\rm yr}^{-1}$.
    \item Main sequence TDE rates should robustly cut off at the main sequence Hills mass, which, depending on the SMBH spin distribution, lies in a range $10^8 \lesssim M_\bullet/M_\odot \lesssim 10^9$.
\end{enumerate}
Conclusion (3) exists in tension with a number of the lower empirical TDE rate estimates discussed in \S \ref{sec:observationalRates}, such as those of \citet{Donley+02}, \citet{KhabibullinSazonov14}, and \citet{vanVelzenFarrar14}.  Another interesting, observationally motivated puzzle is the large rate enhancement seen in post-starburst galaxies (\S \ref{sec:observationalRates}), which was not directly predicted by any pre-existing dynamical studies.

It seems, therefore, that potential tensions exist between the inferred TDE rates and those predicted from (empirically calibrated) dynamical theory. These tensions can be divided into two types: too many TDEs observed in post-starburst galaxies, and too few observed in ``normal'' galaxies. In the following subsections, we discuss both of these tensions. First, however, we must note one important caveat in the comparison of theory to observation: the necessity of having observed TDE samples that are both flux-complete and pure (i.e. not contaminated by a significant number of TDE impostors, such as nuclear supernovae, AGN variability, or more exotic transients).  A large presence of TDE impostors in an observed sample will skew inferred event rates upwards, while a flux-incomplete TDE sample will skew inferred rates downwards.  The first of these issues is discussed at length in the \impostchap, while we cover the latter in some detail below.

\subsubsection{Elevated TDE rates in Post-Starburst Galaxies}
\label{sec:poststarburst}

The post-starburst preference of optical \citep{Arcavi+14} and X-ray \citep{Graur+18} TDEs is an interesting observational puzzle that was not predicted by theory.  
The tension between dynamically predicted rates and the higher rates inferred from observations is a problem that has many possible solutions.  The simple (spherically symmetric, quasi-isotropic) models of \S \ref{sec:applied} neglect many proposed dynamical mechanisms for increasing per-galaxy TDE rates above $\dot{N} \sim 10^{-4}~{\rm yr}^{-1}$, such as binary SMBHs \citep{Ivanov+05, Chen+11}, radial anisotropies \citep[$b>0$;][]{Stone+18}, nuclear triaxiality \citep{Merritt&Poon04}, secular instabilities in stellar discs \citep{Madigan+18}, overdense central star clusters \citep{StoneMetzger16, StonevanVelzen16}, and rate enhancements from massive perturbers \citep{Perets+07a, Perets:2008,Mastrobuono+2014} and/or nuclear spiral arms \citep{Hamers:2017}.  In this subsection, we discuss each of these potential solutions to the discrepancy.

The first proposed explanation for the post-starburst preference invoked the correlation between starbursts and galaxy mergers \citep{Arcavi+14}.  If many post-starburst galaxies are also post-merger galaxies, their nuclei may contain SMBH binaries which can increase TDE rates by many orders of magnitude (relative to galactic nuclei with solitary SMBHs) through a combination of Kozai cycles \citep{Ivanov+05} and chaotic three-body scatterings \citep{Chen+11, Wegg&Bode11}.  However, even though SMBH binaries may temporarily enhance TDE rates by multiple orders of magnitude, the short timescales for such enhancements (typically $\sim 10^5~{\rm yr}$, e.g. \citealt{Wegg&Bode11}) may make it challenging for this mechanism to explain the global fraction of all TDEs seen in post-starburst galaxies (see also discussions in \citealt{StoneMetzger16, Saxton+18}). 

Prior to the discovery of the post-starburst preference, \citet{Perets+07a, Perets:2008} suggested that massive perturbers, such as giant molecular clouds, can greatly shorten the two-body relaxation times (since $T_{\rm rel} \propto \langle M_\star^2 \rangle^{-1}$) and enhance TDE rates as a result.  Nuclear spiral arms would have a similar effect \citep{Hamers:2017}.  Such enhanced rates would occur preferentially in gas-rich galaxies and, in particular, in post-merger galaxies, and would last for long timescales. However, because massive perturbers are unlikely to exist at radial scales $r \lesssim r_{\rm crit}$, these processes could possibly enhance TDE rates by a factor of two or less, and are thus unlikely explanations for the extreme, $1-2$ orders of magnitude enhancements inferred for post-starburst galaxies. 

Nuclear starbursts sometimes produce eccentric stellar discs in which secular effects can dramatically increase TDE rates \citep{Madigan+18, WernkeMadigan18}. However, for such secular processes to operate, the nuclear cluster mass should be relatively small, so as not to give rise to mass precession that quenches coherent secular evolution. This may be problematic for this explanation of the post-starburst preference, as most low-mass SMBHs coexist with a sizeable nuclear star cluster.  The more favorable environment of a disc-dominated nuclear stellar population likely exists only for more massive galaxies, which host SMBHs with $M_\bullet \gtrsim 10^8$ M$_\odot$ \citep{Antonini+15} above the Hills mass $M_{\rm H}$ that are unable to tidally disrupt main sequence stars.

If star formation in starbursts is centrally concentrated\footnote{This seems to be indicated by resolved color gradients in nearby E+A galaxies, see e.g. \citet{Pracy+12}.}, then post-\-starbursts may have unusually overdense galactic nuclei. Such overdensities would result in short two-body relaxation times and high TDE rates \citep{StoneMetzger16}. Nevertheless, it is not yet clear whether such extreme density nuclear clusters form in most post-starbursts (most nuclei can not be spatially resolved to assess the stellar density there, although the results of \citealt{StonevanVelzen16} lend tentative support to this hypothesis in one nearby, highly overdense E+A). 

A final class of dynamical explanations relies on asymmetries in stellar position and/or velocity fields.  If nuclear star formation produces highly triaxial potentials \citep{Merritt&Poon04} or radially anisotropic velocity distributions \citep{Stone+18}, these effects can increase TDE rates by an order of magnitude.  However, both of these types of asymmetries tend to ``wash out'' over time due to the effects of two-body relaxation, which may pose a problem in the low-mass galaxies with short relaxation times that host most TDEs.

With so many theoretical explanations to choose from, it is important to find observational tests that can discriminate between different mechanisms.  
The total TDE rates from the mechanisms suggested above are sensitive to a wide variety of assumptions regarding the host galaxies and their nuclei.  
The statistical properties of post-starburst TDE hosts might therefore be a more useful diagnostic.  For example, the observed distribution of post-starburst host masses is biased towards smaller galaxies, in agreement with the overdensity and radial orbit hypotheses, but in notable disagreement with the top-heavy distribution of host masses expected for SMBH binary-triggered TDEs \citep{Stone+18}.  Likewise, the delay time distribution of TDEs in post-starbursts is highly discrepant with that of the SMBH binary explanation, generally compatible with the overdensity hypothesis, and compatible with the radial orbit hypothesis provided fairly extreme\footnote{Such large anisotropies may be vulnerable to the radial orbit instability \citep{Polyachenko&Shukhman81}.} ($b \gtrsim 0.5$) anisotropy parameters are chosen \citep{Stone+18}.  Dynamical modeling of {\it HST} photometry of the nearby E+A galaxy NGC 3156 lends further support to the overdensity explanation, as this galaxy appears to have both an unusually steep central density profile ($\rho(r) \propto r^{-2.25}$, formally in the ``ultrasteep'' regime) with an unusually small SMBH influence radius \citep{StonevanVelzen16}. 

Finally, many of the suggested scenarios -- such as SMBH binaries, eccentric nuclear discs, radial anisotropies and nuclear triaxiality -- produce TDEs predominantly in the full loss-cone regime (or at the very least, with a $\beta$ distribution matching that of the full loss-cone regime).  TDE flares produced via these channels\footnote{A notable exception to this trend is the overdensity scenario; ultrasteep density cusps will produce almost all their TDEs in the empty loss cone regime.} should preferentially have a lower fraction of grazing TDEs than will TDEs arising from stellar two-body relaxation processes.

While the true explanation of the post-starburst preference remains an open question, it seems likely that (i) larger statistical samples of TDE hosts and (ii) dynamical modeling of more nearby post-starburst nuclei will allow rapid progress on this question in the near future.

\begin{figure}
\includegraphics[width=120mm]{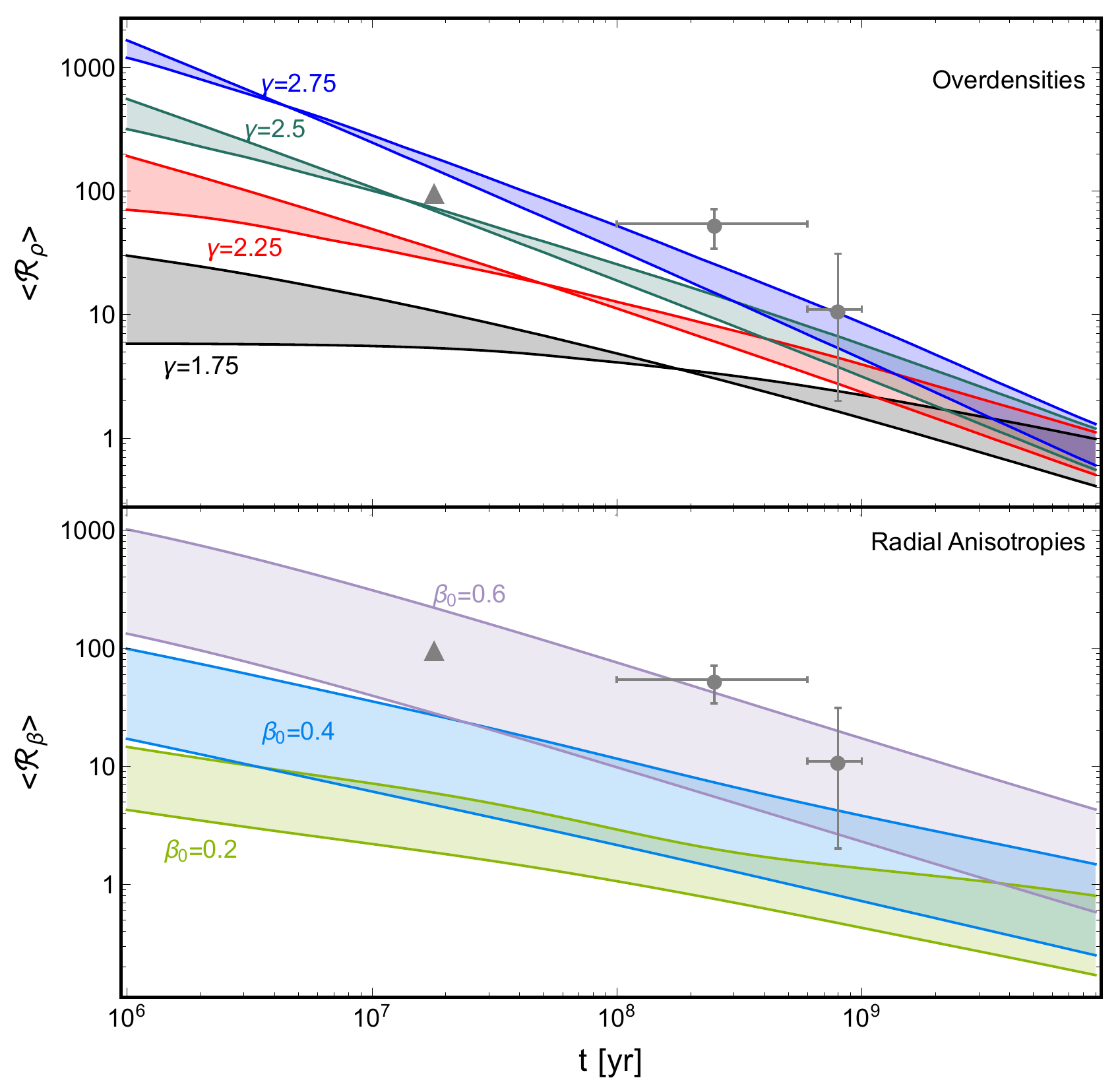}
\caption{Theoretical delay time distributions as functions of time $t$ since a burst of nuclear star formation.  {\it Top panel}: the average rate enhancement $\langle \mathscr{R}_\rho \rangle$ in {\it overdense} galactic nuclei, in comparison to TDE rates from a population of ``normal'' galaxies.  The enhancement factor $\langle \mathscr{R}_\rho \rangle$ is computed by integrating across an empirical SMBH mass function (the boundaries of the shaded error regions correspond to assuming that astrophysical SMBH masses cut off either at $10^5 M_\odot$ or $10^6 M_\odot$).  Theoretical models assume a range of different post-starburst density profiles $\rho(r) \propto r^{-\gamma}$, and circular data points are taken from a sample of eight TDE candidates with well-characterized host galaxies \citep{French+16, French+17}.  {\it Bottom panel}: same as before, but now considering the average rate enhancement $\langle \mathscr{R}_\beta \rangle$ in {\it radially biased} galactic nuclei, with a range of different initial anisotropy parameters $\beta_0$.  Taken with permission from \citet{Stone+18}.}
\label{fig:DTD}
\end{figure}

\subsubsection{A Possible Dearth of TDEs in ``Normal'' Galaxies}
Very few dynamical mechanisms exist to {\it lower} TDE rates below the conservative floor set by two-body relaxation in a spherical and isotropic nucleus.  The only mechanism that seems clearly able to do this is the presence of a strong tangential anisotropy in the stellar velocity field, which is capable of producing arbitrarily large reductions in $\dot{N}$ \citep[$b<0$;][]{Merritt&Wang05}.  However, such tangential anisotropies will wash out due to two-body relaxation on a small fraction of the energy relaxation time \citep{Lezhnin&Vasiliev15}.  The small galaxies that dominate the volumetric TDE rate (and whose relaxation times are less than $t_{\rm H}$), are therefore unlikley to have large tangential anisotropies in their nuclei unless an exotic dynamical mechanism exists to pump $b<0$ in small galactic nuclei.  Long-lived SMBH binaries are expected to produce this type of tangential anisotropy \citep{Merritt&Wang05}, but as small galactic nuclei are the environments most favorable for solving the final parsec problem through collisional mechanisms \citep{Begelman+80}, this does not seem promising.

\begin{figure}
\includegraphics[width=120mm]{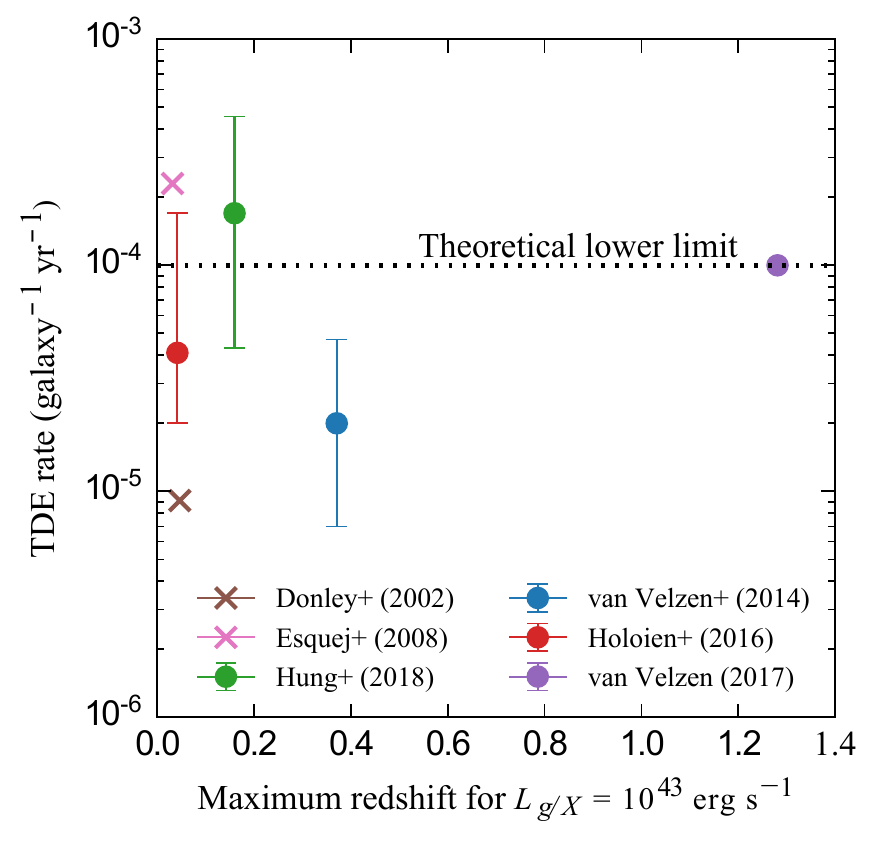}
\caption{Observationally inferred TDE rates from six different (but, in some cases, overlapping) TDE samples: \citet[][brown]{Donley+02}, \citet[][pink]{Esquej+08}, \citet[][blue]{vanVelzenFarrar14}, \citet[][red]{Holoien+16}, \citet[][purple]{vanVelzen18}, and \citet[][green]{Hung+18}.  Rate estimates using TDEs selected by their soft X-ray emission are indicated with ``X'' symbols, while rate estimates using optically-selected TDEs are shown with circles.  The x-axis indicates the maximum redshift $z$ out to which the underlying surveys would detect flares with peak g-band luminosities of $L_{\rm g}=10^{43}~{\rm erg~s}^{-1}$ (for the X-ray samples, the equivalent condition is a 0.2-2.4 keV luminosity $L_{\rm X}=10^{43}~{\rm erg~s}^{-1}$).  The dotted black line shows an approximate theoretical lower limit on per-galaxy TDE rates due to 2-body relaxation in isotropic galactic nuclei \citep{WangMerritt04}.  Taken with permission from \citet{Hung+18}.}
\label{fig:HungComparison}
\end{figure}

The lack of clear mechanisms available to decrease {\it dynamically-predicted} TDE rates led \citet{StoneMetzger16} to identify a ``rate discrepancy'' between (empirically-calibrated) theory and observation\footnote{Although it is worth noting that some observational rate inferences, such as \citealt{Esquej+08}, would not be in tension with conservative theoretical rate estimates.}.  Subsequent analysis of observed TDEs has identified a plausible resolution to this discrepancy: a very steep TDE luminosity function.  In particular, \citet{vanVelzen18} find that, for optically-selected TDE candidates, ${\rm d}\dot{N}/{\rm d}L \propto L^{-2.5}$, with an uncertain lower luminosity limit.  This suggests that current, flux-limited TDE samples are seeing only ``the tip of the iceberg,'' in comparison to the much larger number of very dim TDEs that exist in the local Universe.  While it is important to test the shape of the TDE luminosity function further, both with larger near-future TDE samples {\it and} with better first-principles modeling of flare emission mechanisms, this seems like a promising resolution of tensions between theoretical and observational rate estimates.  The origin of the luminosity function is theoretically uncertain, and depends on the complicated physics of debris circularization (\flowchap{}) and optical emission (\emischap).  Specific hypotheses could be tested with larger TDE samples; for example, rapid disc formation is probably disfavored for $\beta\approx 1$ TDEs around small SMBHs \citep{Dai+15}, so the smaller SMBHs that dominate volumetric event rates $\dot{n}$ might preferentially be associated with lower peak luminosities.   We illustrate the current state of this rate discrepancy in Fig. \ref{fig:HungComparison}, which shows that the most recent rate inferences from optically-selected TDE samples appear compatible with the conservative side of theoretical rate estimates.

\section{Broader Implications}
\label{sec:implications}
In the final section of this Chapter, we focus on the broader scientific importance of TDE rates.  We have already explored one major astrophysical motivation for studying statistical samples of TDEs: SMBH demographic measurements.  In \S \ref{sec:GR}, we saw that the statistical distribution of SMBH spins is imprinted into the mass distribution of TDE host galaxies (at least for $10^{7.5} \lesssim M_\bullet / M_\odot \lesssim 10^{8.5}$).  Likewise, in \S \ref{sec:empirical}, we saw that the volumetric TDE rate contains information on the uncertain bottom end of the SMBH mass function.  Because these motivations have already been described in depth, we will not belabor them in this section, but instead focus on alternative applications and implications of TDE rates.  In \S \ref{sec:demographics}, we discuss the potential importance of TDEs for SMBH mass, spin, and luminosity evolution.  Next, in \S \ref{sec:Hills}, we draw on the general relativistic analysis of \ref{sec:GR} to explore how SMBH spin distributions may be encoded in volumetric disruption rates.  In \S \ref{sec:EMRIs}, we explore how TDE rates and inferences may be used to calibrate our understanding of extreme mass ratio inspirals, a related byproduct of loss cone physics in galactic nuclei.  Finally, in \S \ref{sec:future}, we offer a few speculations for what the future may hold for tidal disruption science.

\subsection{Black Hole Demographics and Growth}
\label{sec:demographics}

As SMBHs feast on debris from tidally disrupted stars, they accrete both mass and angular momentum. 
Over a Hubble time, $t_{\rm H}\approx 1.4\times 10^{10}~{\rm yr}$, the cumulative effect of many TDEs is to increase the SMBH mass, $M_\bullet$.  The net effect on SMBH angular momentum, $\chi_\bullet$, is less obvious, as individual TDEs may either spin it up or spin it down.  Because TDEs usually produce transient accretion flares, the stochastic background of TDEs in otherwise quiescent (inactive) galactic nuclei produces a baseline level of accretion luminosity that may contribute in an interesting way to the bottom end of the AGN luminosity function.

The importance of all three of these effects -- SMBH mass growth, SMBH spin evolution, and accretion luminosity in quiescent galaxies -- depends strongly on TDE rates.  These effects will be of little importance in galaxies with very low TDE rates, but can be of crucial importance in galaxies, or classes of galaxies, with higher TDE rates.  Indeed, it is possible that some types of galaxies may accumulate most of their SMBH mass and/or spin angular momentum through stellar tidal disruption.  

In the simplest picture, where loss cone repopulation is governed by two-body relaxation, TDE rates are decreasing functions of SMBH mass (\S \ref{sec:applied}), indicating that TDEs will be of the greatest importance for the growth of low-mass SMBHs \citep{Milosavljevic+06}.  We may compute the typical SMBH mass below which TDEs are important for growth\footnote{In this calculation, we have assumed that half of the disrupted star accretes onto the SMBH.  For nearly-parabolic stellar orbits, precisely half of the disrupted star is dynamically bound to the SMBH, although we caution that hydrodynamic shocks and radiation pressure in super-Eddington accretion may unbind a portion of this dynamically bound half (see the \flowchap{} and the \diskchap{} for more discussion of these uncertainties).} by equating $M_\bullet = \frac{1}{2}\langle M_\star \rangle \dot{N}(M_\bullet) t_{\rm H}$.  
We solve this equation using the empirically calibrated scaling of Eq. \ref{eq:empiricalRate}, and find $M_{\rm TDE} \approx 5\times 10^5 M_\odot$ to be the characteristic mass below which SMBH growth should be dominated by TDEs.  Larger SMBHs acquire most of their mass from other sources, primarily radiatively efficient gas accretion \citep{Soltan82}.

Because two-body relaxation delivers stars to the SMBH from a quasi-isotropic distribution of directions\footnote{In principle, if TDE rates are dominated by mechanisms (such as nuclear triaxiality, or eccentric stellar discs) that preferentially supply stars from a specific orbital orientation, TDEs may act to spin up SMBHs.}, the net effect of mass growth through TDEs is to spin the SMBH down, in analogy to ``chaotic accretion'' of gas clouds by AGN \citep{King&Pringle06}, and small SMBHs that have grown primarily through tidal disruption should be spinning slowly. 
Interestingly, even if TDEs are a subdominant contributor to SMBH mass growth, they may still play an important role in the evolution of SMBH spin.  SMBHs that grow through prolonged AGN episodes, or through short episodes with the same direction of disc angular momentum 
will, over a handful of mass-doubling times, spin up to very high values of $\chi_\bullet$ \citep{2008ApJ...684..822B}.  If it is only these AGN episodes that supply angular momentum to the SMBH, $\chi_\bullet$ will saturate at a value dictated by accretion physics - for example, growth through standard thin disc accretion saturates at $\chi_\bullet=0.998$ \citep{Thorne74}.  Because this dimensionless spin is so large, even a very small fraction of mass growth from (isotropically distributed) TDEs will place a lower cap on SMBH spin \citep{Metzger&Stone16}.

Even though TDEs are rare events, they occur in all galaxies with $M_\bullet < M_{\rm H}$, and therefore produce a minimum time-averaged accretion rate onto SMBHs in quiescent galactic nuclei.  While the peak luminosity of a TDE (up to $\sim 10^{44-45}~{\rm erg~s}^{-1}$) is typically maintained for a period of weeks to months, there is a long, slow decay period in which the galactic nucleus may resemble a low-luminosity AGN.  Recent UV observations indicate that most TDEs continue to radiate at $\sim 10^{42}~{\rm erg~s}^{-1}$ at times $\sim 10~{\rm yr}$ post-peak \citep{vanVelzen+18}.  Based on simple models for TDE light curves, \citet{Milosavljevic+06} estimated that main sequence TDEs could be responsible for a majority of the lowest-luminosity X-ray AGN (although it is important to note that the ``AGN'' produced by late stages of TDEs would probably lack standard narrow-line regions).  More recent work suggests that episodic partial disruptions of giant-branch stars may be the most important type of tidal disruption for setting a floor on accretion rates in galactic nuclei \citep{MacLeod+13}.  The actual luminosity floor set by the late stages of tidal disruption flares remains somewhat ambiguous, as it depends on (i) the still uncertain volumetric TDE rate, (ii) the presence \citep{Shen&Matzner14} or absence \citep{vanVelzen+18} of thermal instability in late-time TDE discs, and (iii) the ability of TDE debris from red giant disruptions to efficiently accrete onto the SMBH \citep{Bonnerot+16}.

\subsection{The Shadow of the Horizon}
\label{sec:Hills}
As we have seen in \S \ref{sec:GR}, the Hills mass depends sensitively on SMBH spin, and can vary by almost an order of magnitude (for prograde equatorial orbits) as one varies $\chi_\bullet$ from $0$ to $1$.  While this has exciting astrophysical applications for measuring the distribution of SMBH spins in a mass range $10^{7.5} \lesssim M_\bullet/M_\odot \lesssim 10^{8.5}$, the hard upper limit on $M_{\rm H}$ (for lower main sequence stars) imposed by the Kerr bound ($\chi_\bullet \le 1$) raises the prospect of using TDE distributions as probes of general relativity.  For example, \citet{Lu+17} demonstrate that samples of TDEs may rule out exotic alternatives to Kerr black holes, such as boson stars, which possess hard surfaces and would therefore be capable of producing limited electromagnetic emission from stellar impacts onto central massive objects with super-Hills masses.  More generally, we may say that the hypothetical discovery of TDE flares from galaxies with SMBH masses $M_\bullet \gtrsim 10^9 M_\odot$ by future time-domain surveys would strongly motivate consideration of exotic SMBH alternatives.

It is interesting to note that even in a limited sample of twelve optically-selected TDE candidates, there is already statistical evidence for a super-exponential cutoff in the TDE rate near $M_\bullet \sim 10^8 M_\odot$ \citep{vanVelzen18} as discussed in Sec.~\ref{sec:GR}.  This tentative evidence for the Hills mass may provide support for the the identification of observed TDE candidates as bona-fide stellar tidal disruption events; it likewise demonstrates how much larger near-future TDE samples may statistically test the existence of event horizons.

\subsection{EMRI Rates}
\label{sec:EMRIs}

If a star approaching the SMBH is compact enough, the differential gravitational pull exerted on points which are diametrically separated will not lead to dangerously large tidal stresses. A compact object such as a neutron star, a white dwarf, or a stellar-mass black hole can therefore safely pass through pericentres with $R_{\rm p} \gtrsim R_{\rm mb}$\footnote{The marginally bound radius $R_{\rm mb}$ is the minimum pericentre that avoids capture by the event horizon and is of order $R_{\rm g}$ \citep{1972ApJ...178..347B}.} many times until it plunges through the SMBH event horizon due to energy loss via gravitational radiation. This chain of events is referred to as an ``extreme mass-ratio inspiral'' (EMRI, see the review of \citealt{2018LRR....21....4A} and references therein). The number of passages is roughly proportional to the mass ratio, so that for a $10\,M_{\odot}$ stellar-mass black hole and a SMBH of mass $10^6\,M_{\odot}$, the EMRI will complete some $\sim 10^5$ orbits before plunging.

This source of gravitational waves represents a unique probe of gravity in the strong-field regime.  Generic EMRI orbits are both eccentric and inclined, and apsidal and nodal precession imply that the compact object explores the full torus between the radial and polar turning points as it inspirals towards the black hole.
At every pericentre passage there is a strong burst of radiation, and
with a typical total of $\sim 10^5$ such bursts, an EMRI can be used to map
spacetime around SMBHs. With EMRIs one can probe the geometry of SMBHs 
in a regime of non-linear, strong-field and dynamical gravity. This
means that one can test GR and alternative theories of gravity, but also test
different questions related to stellar dynamics in galactic nuclei.  The gravitational-wave
radiation from these systems allows one to extract qualitative information from
a regime which is inaccessible to the photon, the surface of the central
supermassive dark object, and to investigate if spacetime around it differs
from Kerr.  In particular, in GR the Kerr solution is the unique end state of
gravitational collapse. The Kerr metric depends on the mass and spin of the
object only, and all higher multiple moments are related to the gravitational
radiation emitted by the compact object on its orbits towards the SMBH. This
can be used to test the no-hair theorem, as first put forward by
\cite{1995PhRvD..52.5707R}.  Moreover, they can be used as standard sirens to
measure the luminosity of the source and to test the cosmological expansion
history of the Universe \citep{1986Natur.323..310S}. See \cite{2007CQGra..24R.113A,2015JPhCS.610a2002A} for
a general description of the implications of EMRIs on cosmology and fundamental
physics.

EMRIs are formed via two-body relaxation in dense stellar systems, and the theory that allows us to derive their event rate is heavily based on the concept of a loss cone. This is so because of the many similarities of the two problems. However, as we mentioned before, the compact object needs to revolve for tens or hundreds of thousands of orbits around the SMBH before disappearing into the horizon, not just one orbit, as is the case for TDEs. 
When an orbit's pericentre approaches $R_{\rm p}\sim {\rm few}\times R_{\rm g}$, even if its semimajor axis is $a\gg R_{\rm g}$, it starts to rapidly lose energy and progress towards smaller $a$ while remaining at a nearly constant $R_{\rm p}$, until it is finally captured at $a\sim {\rm few}\times R_{\rm g}$ (see e.g. Fig.~43 of \citealt{2018LRR....21....4A}). However, for this process to be successful, the inspiralling object should not be disturbed by the two-body relaxation. This places an upper limit on the initial values of $a$ for EMRIs at around $10^{-2}$~pc, meaning that EMRIs constitute a small fraction of all compact object capture events ($\lesssim 1\%$, see Fig.~17 in \citealt{BarOr&Alexander16}).
The accumulation of a long number of orbits means that relativistic effects at pericentre must be taken into account.

Analogous but even stricter considerations limit the parameter space for ``stellar EMRIs,'' which are main-sequence stars inspiraling on quasi-circular orbits.  These systems are potentially interesting because of their long lives: stable Roche lobe overflow can last for millions of years, and power a weak accretion disk \citep{Dai&Blandford13}, though the arrival of a second stellar EMRI during the lifetime of the first may lead to a TDE impostor due to mass loss in a hypersonic stellar collision \citep{Metzger&Stone17}.  However, such systems require fairly fine-tuned initial conditions in order to circularize through gravitational wave emission (as opposed to tidal circularization, which would destroy the star), and the most promising formation scenario may use the bound star left over after the Hills mechanism breaks up a pre-existing binary \citep{AmaroSeoane+12}.

Unless the SMBH is Schwarzschild, the spin plays a fundamental role in the fate
of a compact object EMRI \citep{2013MNRAS.429.3155A}, controlling the required number of pericentre passages 
before crossing the event horizon. Resonant relaxation can also
allow EMRIs to form closer to the SMBH, where the stellar density decreases,
and two-body relaxation is not efficient enough
\citep{2006ApJ...645.1152H}. 
While the dynamics of resonant relaxation are much more complicated than that of two-body relaxation, recent studies indicate that its ultimate contribution to steady-state EMRI rates is modest \citep{Merritt15d, BarOr&Alexander16, 2017JPhCS.840a2019A}.

Even though a consensus about the cusp at the Galactic Centre has recently
emerged in which theory, numerical calculations and observations agree on the
existence of the sub-pc stellar cusp,
\citep{2018A&A...609A..26G,2018A&A...609A..27S,2018A&A...609A..28B}, the dynamics of distant galactic nuclei on sub-pc scales are less clear, making it challenging to predict generic EMRI rates.  The first major astrophysical uncertainty concerns the massive black hole occupation fraction.  If many intermediate-mass black holes exist in the nuclei of dwarf galaxies, their contribution to the volumetric EMRI rate could dominate over larger SMBHs \citep{Babak+17}.  A second major uncertainty comes from the unknown distribution of stellar profiles; if   
cored galactic nuclei are more common than is expected, volumetric EMRI rates will be low. Finally, as noted
before, the magnitude of SMBH spin is crucial in the derivation of
EMRI rates \citep{2013MNRAS.429.3155A}, as rapidly spinning Kerr SMBHs will convert many ``direct plunges'' into observable EMRIs. 
Upcoming observations of TDEs in nuclei which harbour SMBHs in the mass range of $10^5-10^7\,M_{\odot}$ may help to at least partially clear up these uncertainties; using large samples of TDEs to constrain the low-mass black hole occupation fraction will be particularly valuable.

\subsection{Future Prospects}
\label{sec:future}

Over the next five years, our sample of TDEs will expand by multiple orders of magnitude.  At the time of writing, a few dozen strong TDE candidates have been discovered, primarily through thermal emission in either the soft X-ray (e.g. the compilations of \citealt{Komossa15} and \citealt{Auchettl+17}, and the \xraychap{}) or optical/UV (e.g. the compilation of \citealt{Hung+17} and the \optchap{}) bands\footnote{The online ``Open TDE Catalog'' https://tde.space/ is a useful resource for the observationally-curious reader.}.  For most of the last decade, the time-averaged TDE detection rate has been a handful of strong candidate flares per year.  This situation is already starting to change in 2019, as the wide-field optical survey ZTF \citep{Bellm+19} has begun harvesting TDE candidates, with an expected detection rate of $32^{+41}_{-25}$ TDEs per year \citep{Hung+18}.

While the ongoing ZTF survey is expected to increase today's TDE sample by one order of magnitude, near-future optical and X-ray time domain surveys may increase our TDE inventory even further, into the thousands and possibly tens of thousands.  Two upcoming space-based X-ray surveys, {\it eROSITA} \citep[][launched during the writing of this Chapter]{Merloni+12} and {\it Einstein Probe} \citep[][launch date note yet determined]{Yuan+15}, hold particular promise.  The {\it Einstein Probe} is expected to detect tens to hundreds of X-ray bright TDEs per year \citep{Yuan+15}, while {\it eROSITA} predictions give a per-year detection rate $\approx 1000$ \citep{Khabibullin+14}.  On the ground, the wide-field LSST optical survey \citep{Abell+09} has long been recognized for its ability to detect thousands of TDEs per year \citep{StrubbeQuataert09, vanVelzen+11}; the true detection rate depends on survey strategy, but one recent estimate predicted $\approx 3-6 \times 10^3$ TDEs per year.

It is important to consider the many science goals that can be accomplished with such large, near-future samples of TDEs.  Advance planning is important both (i) to develop relevant theoretical predictions that can be tested and falsified by statistical samples of TDEs, and (ii) to aid future observers in prioritizing scarce spectroscopic or multiwavelength followup resources, which will certainly be inadequate to fully characterize {\it every} TDE candidate found by the surveys described above.  In the list below, we offer a brief (and, unavoidably, subjective) selection of important questions in physics and astrophysics that may be partially addressed by large samples of TDEs.  In assembling this list, we have purposefully avoided questions focused entirely on the process and consequences of stellar tidal disruption.  While this arena features many interesting open questions, they are discussed in much greater detail in the other Chapters of this Volume, and here our aim is to examine the utility of TDEs {\it as a tool} for answering broader questions.
\begin{itemize}
    \item What is the distribution of SMBH masses in the Universe?
    \item How did SMBH seeds form at high redshift?
    \item What is the distribution of SMBH spins in the Universe?
    \item Are the ``SMBHs'' observed in galactic nuclei true Kerr metric black holes, or more exotic, horizonless compact objects?
    \item What is the rate of other processes involving loss cone physics in galactic nuclei (e.g. hypervelocity star and EMRI production)?
    \item What fraction of galactic nuclei host unresolved binary SMBHs?
    \item What exotic dynamical processes are common to post-starburst galactic nuclei?
    \item Are there other subclasses of galaxies with anomalous nuclear stellar dynamics?
\end{itemize}
In this Chapter, we have discussed how each of these questions may be probed -- at least to some extent -- with  a statistical sample of TDEs.  But in almost all cases, directly answering these questions with TDEs will require a firmer grasp on rates of tidal disruption, and how they depend on astrophysical variables like SMBH mass, host galaxy mass, redshift, and so on.  Over the next few years, we expect progress on TDE rates to arise from three parallel channels.  Better theoretical models for non-standard loss cone repopulation mechanisms (e.g. nuclear triaxiality, eccentric stellar discs) will help solidify our understanding of underlying dynamical theory.  Better dynamical modeling of nearby galactic nuclei will improve our understanding of TDE rates in astrophysically realistic galaxies.  And better samples of observed TDE flares -- samples that are both larger and more homogeneously selected -- will provide firmer observational estimates of the real volumetric TDE rate in the Universe.

\begin{acknowledgements}
N.C.S. received financial support from the NASA Astrophysics Theory Research Program (Grant NNX17AK43G; PI B. Metzger).  M. K. acknowledges support from NSF Grant No. PHY-1607031 and NASA Award No. 80NSSC18K0639.  E.M.R. acknowledges support from NWO TOP grant Module 2, project number 614.001.401.  P.A.S. acknowledges support from the Ram{\'o}n y Cajal Programme of the Ministry of Economy, Industry and Competitiveness of Spain, the COST Action
GWverse CA16104, and the National Key R\&D Program of China (2016YFA0400702) and the National Science Foundation of China (11721303).
\end{acknowledgements}

\bibliography{main}
\bibliographystyle{aps-nameyear}          

\end{document}